%% file: main.tex
\documentclass[lettersize, journal]{IEEEtran}
\IEEEoverridecommandlockouts
\usepackage{cite}
\usepackage{amsmath,amssymb,amsfonts}
\usepackage{algorithmic}
\usepackage{graphicx}
\usepackage{textcomp}
\usepackage[dvipsnames]{xcolor}
\usepackage{url}
\usepackage{tikz}
\usepackage{todonotes}
\usepackage{float}
\usepackage{subfigure}
\usepackage{siunitx}
\DeclareSIUnit{\gbps}{Gb/s}
\DeclareSIUnit{\gb}{Gb}

\usepackage{hyperref}
\hypersetup{
    colorlinks = true,
    allcolors=blue,
    urlcolor=black,
    linkcolor=black
}
\usepackage[]{acronym}
\newcommand*\ballnumber[1]{\tikz[baseline=(char.base)]{
            \node[shape=circle,fill,inner sep=.5pt] (char) {\textcolor{white}{#1}};}}

\newcommand{\SL}[1]{{#1}}

\usepackage{fancyhdr}
\fancypagestyle{IEEEtitlepagestyle}{
    \fancyhf{} 
     
    \fancyhead[C]{
        \small © 2024 IEEE. Personal use of this material is permitted. Permission from IEEE must be obtained for all other uses, including reprinting/republishing this material for advertising or promotional purposes, collecting new collected works for resale or redistribution to servers or lists, or reuse of any copyrighted component of this work in other works.\\
        \small \textit{Digital Object Identifier 10.1109/TNSM.2024.3434337}
        }
}

\input{macros}

\begin{document}

\title{P4-PSFP: P4-Based Per-Stream Filtering and Policing for Time-Sensitive Networking\\}
\author{\IEEEauthorblockN{
        Fabian~Ihle,
		Steffen~Lindner,
		and Michael~Menth}
		
	\IEEEauthorblockA{
		University~of~Tuebingen,
		Chair~of~Communication~Networks,
		72076~Tuebingen,
		Germany\\
	}
	\IEEEauthorblockA{
        Email: 
		\{%
        fabian.ihle,~%
		steffen.lindner,~%
		menth\}@uni-tuebingen.de
    }
      \thanks{The authors acknowledge the funding by the Deutsche Forschungsgemeinschaft (DFG) under grant ME2727/3-1. 
    The authors alone are responsible for the content of the paper.}
}

\maketitle

\begin{abstract}
\acf{TSN} extends Ethernet to enable real-time communication.
\revieweri{In \acs{TSN}, bounded latency and zero congestion-based packet loss are achieved through mechanisms such as the \acf{CBS} for bandwidth shaping and the \acf{TAS} for traffic scheduling.}
Generally, TSN requires streams to be explicitly admitted before being transmitted.
To ensure that admitted traffic conforms with the traffic descriptors indicated for admission control, \acf{PSFP} has been defined. 
For \FIii{credit-based metering}, well-known token bucket policers are applied. 
However, \FIii{time-based metering} requires time-dependent switch behavior and time synchronization with \FIii{sub-microsecond} precision. 
While TSN-capable switches support various TSN traffic shaping mechanisms, \FIi{a full} implementation of PSFP is still \FIi{not available}.
To bridge this gap, we present a P4-based implementation of PSFP on a 100 Gb/s per port hardware switch.
We explain the most interesting aspects of the PSFP implementation whose code is available on GitHub\footnote{\url{https://github.com/uni-tue-kn/P4-PSFP}}.
We demonstrate credit-based and time-based policing and \FIii{synchronization capabilities} to validate the functionality and effectiveness of P4-PSFP.
The implementation scales up to 35840 streams depending on the stream identification method.
P4-PSFP can be used in practice as long as appropriate TSN switches lack this function. 
Moreover, its implementation may be helpful for other P4-based hardware implementations that require time synchronization.
\end{abstract}


\input{chapters/01-introduction}
\input{chapters/02-related-work}
\input{chapters/03-psfp}
\input{chapters/04-p4}
\input{chapters/05-implementation}
\input{chapters/06-evaluation}
\input{chapters/065-discussion}
\input{chapters/07-conclusion}
\input{glossar}

\bibliography{conferences, literature} 
\bibliographystyle{ieeetr}

\begin{IEEEbiography}
	[{\includegraphics[width=1in,height=1.25in,clip,keepaspectratio] 
		{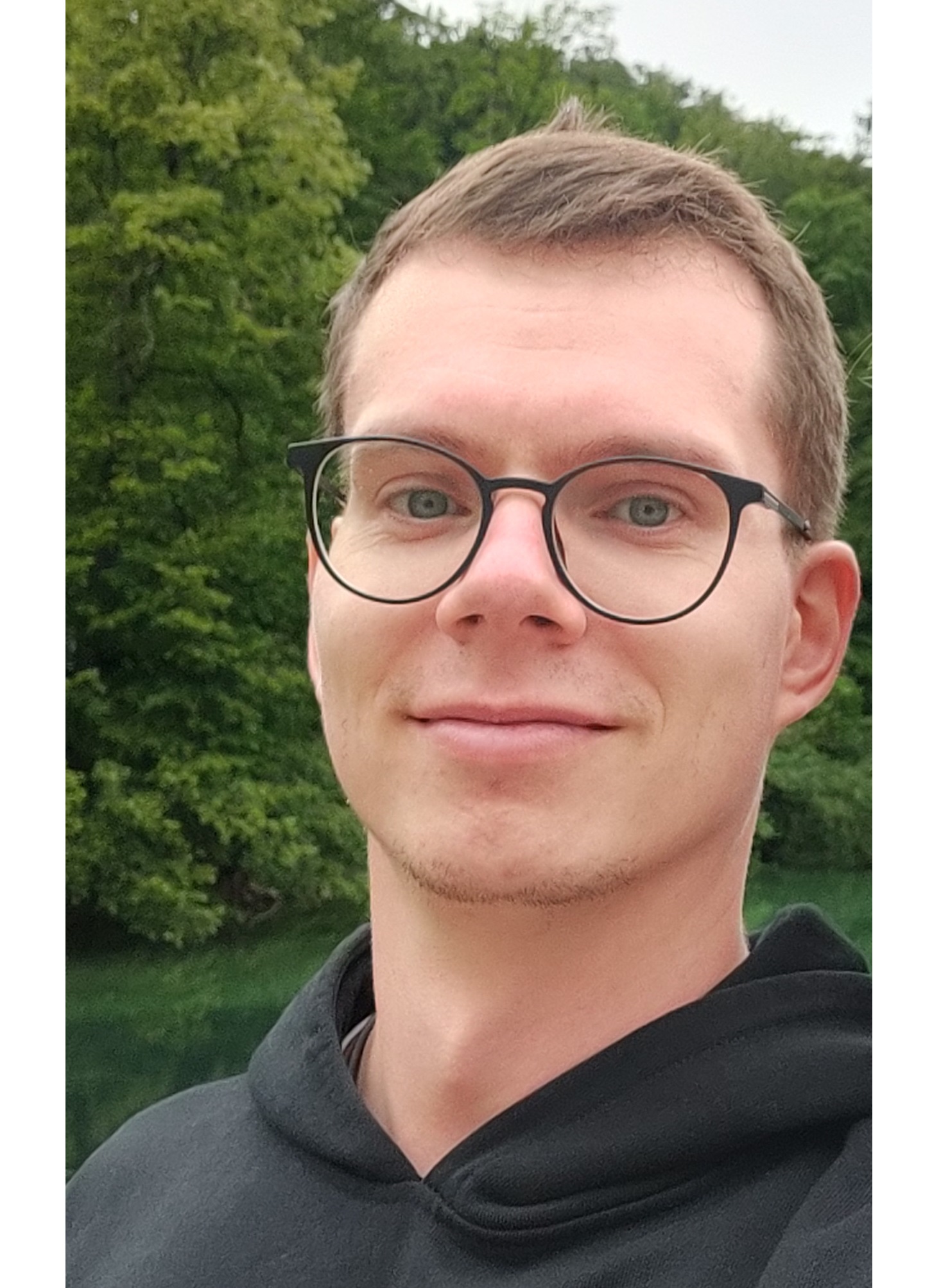}}]{Fabian Ihle }
    received his bachelor's (2021) and master's degrees (2023) in computer science at the University of Tuebingen. Afterwards, he joined the communication networks research group of Prof. Dr. habil. Michael Menth as a Ph.D. student.
    His research interests include software-defined networking, P4-based data plane programming, resilience, and Time-Sensitive Networking (TSN).
\end{IEEEbiography}

\begin{IEEEbiography}
    [{\includegraphics[width=1in,height=1.25in,clip,keepaspectratio] 
        {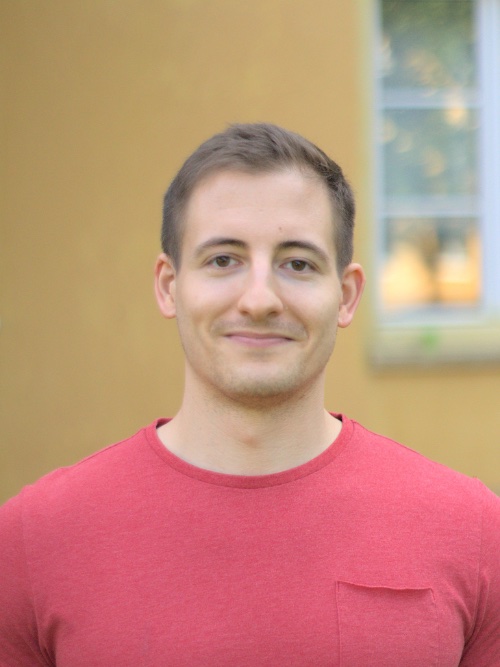}}]{Steffen Lindner }
    is a postdoctoral researcher specialized in software-defined networking (SDN), P4, time-sensitive networking (TSN), and congestion management. He studied, worked, and obtained his bachelor’s (2017), master’s (2019), and Ph.D. (2024) degrees at the University of Tuebingen.
\end{IEEEbiography}

\begin{IEEEbiography}
	[{\includegraphics[width=1in,height=1.25in,clip,keepaspectratio]
		{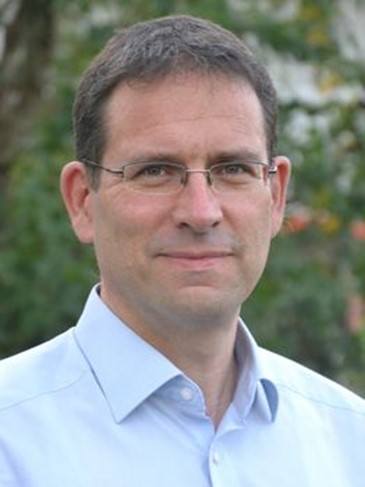}}]{Michael Menth, }
	(Senior Member, IEEE) is professor at the Department
of Computer Science at the University of Tuebingen/Germany and
chairholder of Communication Networks since 2010. He studied,
worked, and obtained diploma (1998), PhD (2004), and habilitation
(2010) degrees at the universities of Austin/Texas, Ulm/Germany,
and Wuerzburg/Germany. His special interests are performance
analysis and optimization of communication networks, resilience and
routing issues, as well as resource and congestion management. His
recent research focus is on network softwarization, in particular
P4-based data plane programming, Time-Sensitive Networking (TSN),
Internet of Things, and Internet protocols. Dr. Menth contributes
to standardization bodies, notably to the IETF.
\end{IEEEbiography}

\vspace{12pt}
\color{red}
\end{document}

%% file: macros.tex
\newcommand\citeN\cite

\newcommand\fig[1]{Figure~\ref{fig:#1}}

\newcommand\sect[1]{Section~\ref{sec:#1}}
\newcommand\equa[1]{Equation~\ref{eq:#1}}
\newcommand\tabl[1]{Table~\ref{tab:#1}}

\newcommand{\FI}[1]{{#1}}
\newcommand{\FIi}[1]{\textcolor{black}{#1}}
\newcommand{\FIii}[1]{\textcolor{black}{#1}}
\newcommand{\revieweri}[1]{\textcolor{black}{#1}}
\newcommand{\reviewerii}[1]{\textcolor{black}{#1}}

\newcommand{\MMii}[1]{\textcolor{black}{#1}}

\newcommand{\figeps}[3][]{%
 \begin{figure}[htb!]
  \begin{center}
   \leavevmode
      \parbox[t]{#1}{%
        \resizebox{#1}{!}{\includegraphics{figures/#2}}
      }
   \caption{#3\vspace{-0.2cm}}
   \label{fig:#2}
  \end{center}
 \end{figure}
}

\newcommand{\threesubfigeps}[7]{
  \begin{figure*}[htb]
    \leavevmode
    \begin{center}
     \subfigure[#2]{
        \label{fig:#1}
        \parbox[t]{0.31\textwidth}{%
            \resizebox{0.30\textwidth}{!}{\includegraphics{figures/#1}}
     \vspace{-1cm}
        }
     }
     \subfigure[#4]{
        \label{fig:#3}
        \parbox[t]{0.31\textwidth}{%
            \resizebox{0.30\textwidth}{!}{\includegraphics{figures/#3}}
     \vspace{-1cm}
        }
     }
     \subfigure[#6]{
        \label{fig:#5}
        \parbox[t]{0.31\textwidth}{%
            \resizebox{0.30\textwidth}{!}{\includegraphics{figures/#5}}
        }
     }
    \end{center}
    \caption{#7}
  \end{figure*}
}

\newboolean{makevspace}
\newcommand{\cvspace}[1]{%
    \ifthenelse
        {\boolean{makevspace}}
        {\vspace{#1}}
        {}%
    }

%% file: chapters/01-introduction.tex
\section{Introduction}
Ethernet-based communication has found its way into industrial \FI{applications} \FIi{such as Industry 4.0, the Internet of Things, and in-vehicle networks.
In this environment, time-critical communication is prevalent.
Failure to meet the strict real-time requirements of mission-critical systems can lead to immediate degradation of factory performance, endanger factory personnel, or put traffic participants at risk.
\FIii{However, communication in Ethernet networks using mechanisms such as best-effort transmission, traffic classification, and VLANs does not provide \ac{QoS} that meets real-time requirements\cite{mess, BeMe21}}.}


\acf{TSN} is a collection of standards for enhancing Ethernet communication \FIi{to meet the stringent real-time requirements of industrial and in-vehicle applications}.
\FIi{The collection} provides mechanisms to achieve zero congestion-based \SL{packet} loss and a deterministic upper bound on latency.
To that end, senders, so-called talkers, signal their QoS requirements \SL{and sending behavior} before they start their transmission.
Bridges make resource reservations for approximately periodic flows, called streams, initiated by talkers, \FIi{e.g., via the stream reservation protocol in IEEE Std 802.1Qat\cite{qat}.}
\FI{This is referred to} as admission control.
Many traffic shaping mechanisms exist in the \acs{TSN} standards to ensure that \SL{the previously signaled QoS} requirements are met\SL{, e.g., \FIi{the} \acf{CBS} in IEEE Std 802.1Qav\cite{qav} or \FIi{the} \acf{TAS} in IEEE Std 802.1Qbv\cite{qbv}}.
%
However, in pure \acs{TSN}, no entity enforces a talker's \FIi{compliance} with \SL{its announced sending behavior}. 
Malicious or misconfigured participants can \SL{therefore} consume more resources than \SL{announced} \FIi{for} admission control. 
As a result, the network may fail to meet its guarantees of a deterministic upper bound on latency and zero congestion-based \SL{packet} loss.
\ac{PSFP}\cite{qci} defined in IEEE Std 802.1Qci\cite{qci} is a mechanism that addresses this problem by limiting or excluding out-of-profile traffic.
\acs{PSFP} consists of three components, \SL{called stream filters, stream gates, and flow meters}. Stream filters identify and filter streams, stream gates monitor the transmission time, and flow meters monitor the bandwidth.
Although there are \SL{some} \acs{TSN}-capable switches that support frame prioritization, \acs{CBS}, and \acs{TAS}, to the best of our knowledge, there are no switches that support the \acs{PSFP} mechanism \FI{in a comprehensive way and in compliance with IEEE Std 802.1Qci\cite{qci}.}
With the emergence of data plane programmability, users can specify the behavior of switches by themselves allowing for mechanisms, such as \acs{PSFP}, to be implemented on their own hardware.

The contribution of this paper is as follows. 
First, we develop P4-PSFP, a novel implementation of the IEEE Std 802.1Qci PSFP mechanism on real hardware in the P4 programming language.
The implementation \SL{covers all} PSFP components, \SL{i.e.,} stream filter, stream gate, and flow meter. 
Second, we leverage the capabilities of a hardware-based switching ASIC, specifically an Intel Tofino™, to achieve line rate processing of \SI{100}{\gbps} per port \SL{and nanosecond precision for time-based operations}.
We successfully address challenges when using real hardware, such as the periodicity in time-based metering.
Furthermore, we propose an approach to account for time inaccuracy in implementations \FIi{where the application of time synchronization protocols such as the \acf{PTP} is not possible.}
This approach includes a mechanism to synchronize time-critical stream gates with the network time and to account for clock drift.

We conduct comprehensive evaluations regarding the functionality of the implemented PSFP components, i.e., credit-based and time-based policing.
Furthermore, we demonstrate the effectiveness of \FI{the} implementation in protecting a \acs{TAS} schedule by aiming to eliminate queueing in a congested network environment.
Finally, we evaluate the scalability of \FI{the} implementation regarding the number of supported streams.

The remaining paper is structured as follows.
In \sect{relwork} we provide background information on \acs{TSN} and review related work.
\SL{\sect{psfp} explains the basic concept of the PSFP mechanism followed by an introduction to the P4 programming language in \sect{p4}}.
\FI{The} implementation of the PSFP mechanism in P4 is explained in \sect{implementation}.
\SL{The} implementation is evaluated in \sect{eval}.
\reviewerii{Finally, we discuss requirements of P4-PSFP and its applicability beyond industrial networks in \sect{discussion}} and conclude the paper in \sect{conclusion}.

%% file: chapters/02-related-work.tex
\section{Background \FI{of} TSN \FI{and} Related Work}
\label{sec:relwork}
In this section, we first give background information on \ac{TSN}. 
We then review related work.
\subsection{Time-Sensitive Networking (TSN)}
\label{sec:tsn}
\ac{TSN} is a collection of standards for enhancing Ethernet communication that guarantees QoS requirements, i.e., a deterministic upper bound on latency and zero congestion-based packet loss.
It is currently being standardized by the IEEE 802.1 TSN Task Group.
A \acs{TSN} network consists of several bridges and end stations. 
A sending end station is a talker, a receiving end station is a listener.

A \FIi{data flow} initiated by a talker is called a \acs{TSN} stream and must first be admitted by the network before it can be transmitted.
Admission control implies that a talker signals the properties of a new stream to be admitted, e.g., the frame size and \acs{QoS} requirements, to the network. 
\revieweri{We call the description of these properties a stream descriptor.}
The talker agrees to adhere to its signaled stream properties so that the network can rely on this specification for resource management purposes.
Based on this information, the network accepts the request and reserves resources for the new stream, or it rejects the new stream for high-priority transmission.
\acs{TSN} streams may coexist with best-effort transmissions which do not require admission control.

A \acs{TSN} network can use different methods to guarantee latency bounds, such as the \acf{CBS} in IEEE Std 802.1Qav\cite{qav} that uses \FIii{a token bucket mechanism}, and the \acf{TAS} introduced in IEEE Std 802.1Qbv\cite{qbv}.
The \acs{TAS} divides the communication into repeating cycles composed of time slices, similar to a \ac{TDMA}-based approach\cite{mess}, but for packet-switched communication networks.

\revieweri{
Bridges in IEEE Std 802.1Q\cite{8021q} place frames in a queue according to the priority in the \ac{PCP} field in the VLAN tag of the frame.
Based on the assigned queue, \acs{TSN} traffic shaping mechanisms can be applied to the frame.
By using an IP stream identification function in combination with an active VLAN stream identification from IEEE Std 802.1CB\cite{cb}, a VLAN tag can be pushed onto frames that match a particular stream.
This makes untagged frames eligible for traffic shaping in \acs{TSN} bridges.}

There can be up to eight different priorities \FI{in IEEE Std 802.1Q}.
\FIi{Within a TSN bridge, each egress port can accommodate up to eight egress queues.
Each queue is associated with one of the eight frame priorities specified in the VLAN header of a frame.}

\MMii{In \ac{TSN}, high-priority streams may be scheduled, i.e., their sending times at talkers are coordinated in such a way that their frames experience hardly any delay in \FIii{bridges}.
Scheduled streams are protected from other traffic by the \acf{TAS}. 
The \acs{TAS} defines a gating mechanism for each egress queue on an egress port. 
It can be used to provide transmission resources exclusively to scheduled traffic, which supports ultra-low latency.
The gating mechanism defines time slices in which only configured egress queues can send traffic.
This is referred to \FIii{as} an egress queue with a gate in a closed or \FIii{an} open state within a time slice.
Frames in a queue with an open gate are selected for transmission in a FIFO manner while frames in a queue with a closed gate are not selected and remain in the queue.
The time slices and gates of the queues are configured in the so-called \ac{GCL}. It consists of several time slices, each associated with an 8-bit vector indicating the queues with an open gate. This \ac{GCL} is processed repeatedly, i.e., time slices and gates are periodic.
For stream scheduling, the periodic transmission times at talkers are calculated offline together with the \acp{GCL} of all bridges. The result is called a schedule \cite{OlCr18, qbv}.}



\FIii{Scheduled streams are sensitive to errors such as a lack of time synchronization between talkers and bridges.
If frames are transmitted by non-synchronized talkers outside their assigned time slice, they possibly consume the resources reserved for other streams. 
As a result, these streams and possibly also others may not meet their latency bounds.}

Thus, \acs{TSN} talkers, listeners, and bridges must be time\FI{-}synchronized so that all participants in a TSN network share a common understanding of time.
High-precision time synchronization is essential for talkers to ensure that they transmit their data at the precisely designated times.
At the same time, \FIii{bridges require accurate time synchronization to determine the arrival time of an incoming frame and to assign it to the correct time slice in the \acs{GCL}.}
\FI{In \acs{TSN} networks}, protocols such as \acs{PTP}\cite{ptp} are used which achieve a precision in the order of tens of nanoseconds\cite{dptp}.


\subsection{Related Work}
We first discuss related work on \acs{PSFP} and similar mechanisms, such as \ac{TDMA}, in various applications and review an implementation of a stream reservation enforcement mechanism in P4.
We then review a hardware traffic generator implemented in P4 called P4TG that is used in \sect{eval}.

\subsubsection{PSFP in \FI{D}ifferent \FI{A}pplications}
Meyer \textit{et al.}\cite{MeHä21} analyze \acs{PSFP} as a security mechanism to identify misbehaving traffic \revieweri{streams} in cars.
They show that \acs{PSFP} reliably detects attacks on car communication in their simulated OMNeT++\cite{omnet} environment in several scenarios.
Similarly, Luo \textit{et al.} \cite{LuWa21} propose to use PSFP for the detection of anomalies caused by DoS attacks and abnormal traffic behavior.
Their evaluations indicate that \acs{PSFP} ensures real-time performance in \acs{TSN} networks and successfully blocks DoS attacks.
Their implementation is based on the OMNeT++ simulation framework and is not implemented on real hardware. 

Nsaibi \textit{et al.}\cite{NsLe17} provide a migration from the industry standard Sercos III, a TDMA-based real-time Ethernet protocol\cite{sercos} with a line topology, to \acs{TSN} networks with a tree topology by extending Sercos III with the TSN standards IEEE Std 802.1AS-Rev\cite{as} and IEEE Std 802.1Qbv\cite{qbv}.
They used a \SI{1}{\gbps} TSN switch for this purpose.
\ac{TDMA} as used in Sercos III is a mechanism very similar to stream gates in \acs{PSFP} where frames are transmitted only during specific time slices.
They conclude that \acs{TSN} brings many benefits for Sercos III networks, such as multi-gigabit data rate and the support for different network topologies.

Szancer \textit{et al.}\cite{SzMe18} aim to improve latency and jitter in TSN-migrated Sercos III networks by integrating not only a single \SI{1}{\gbps} TSN switch, but by enabling TSN for every device in the network.
They tested their implementation with a TDMA-based and a CBS-based approach.
Their simulation shows that a TDMA-based approach has a significantly lower jitter.
Since PSFP essentially consists of both TDMA-based and CBS-based approaches, a Sercos III network migrated to a TSN network could benefit from the additional security features of \acs{PSFP}.

Bülbül \textit{et al.} implemented a mechanism called \acs{TSN} gatekeeper to enforce stream reservations in P4\cite{BüKr23}. 
Their implementation contains elements of the PSFP mechanism, such as stream filtering, stream gates, and flow metering.
However, their stream gate implementation \FIi{remains continuously in the open state, i.e., time-based metering is not applied.}
Multiple streams are assigned to a single stream gate and only the aggregated bandwidth of these streams is monitored.
The transmission times of \acs{TSN} talkers are not monitored. 
Unlike the IEEE 802.1Qci standard, their implementation does not use the stream identification functions defined in IEEE Std 802.1CB\cite{cb}. 
They implemented the TSN gatekeeper mechanism in P4 on the BMv2\cite{bmv2} software target platform and emulated their environment in Mininet.
However, BMv2 implementations are not an indicator whether algorithms will also work on real hardware.

In contrast, \FI{the} implementation of \FI{P4-PSFP} in this work supports \FI{credit-based metering}, cyclic schedules, \FI{i.e., time-based metering}, an approach for synchronizing \FI{the} implementation with the network time and accounting for clock drift, multiple stream identification functions, and is implemented on the Intel Tofino™ switching ASIC hardware target platform which offers high-performance packet processing.

\subsubsection{Deterministic Networking (DetNet)}
\label{sec:detnet}
\revieweri{
\acs{TSN} defines mechanisms to achieve bounded latency and zero packet loss at the link layer over standard Ethernet.
Deterministic Networking (DetNet) is a similar approach using explicit bandwidth reservation, priority queueing, and service protection at the network layer.
DetNet is currently being standardized by the IETF.
In RFC 9023\cite{rfc9023}, the authors describe an approach for operating a DetNet data plane with IP over a \acs{TSN} sub-network.
They state that the efforts in \acs{TSN} to achieve bounded latency and zero packet loss are likely compatible with DetNet networks. 
RFC 9037\cite{rfc9037} describes a similar approach carrying traffic over DetNet-capable MPLS nodes interconnected by a \acs{TSN} subnet.
In this approach, the stream ID in the TSN subnet is derived from the MPLS flow parameters.
Therefore, service protection mechanisms such as \acs{PSFP} can be applied to MPLS flows carried over the TSN subnet.}

\revieweri{Choi \textit{et al.}\cite{ChoKa22} describe an implementation of a DetNet packet forwarding engine operating at \SI{100}{\gbps}. 
They can guarantee a bounded latency with less than \SI{6.25}{\micro\second} per node.
Addanki \textit{et al.}\cite{AdIa20} propose a mechanism called Time Aware PIFO Scheduling (TAPS) to address the problem of misbehaving flows in DetNet which is also a problem in \acs{TSN} networks.
The authors of \cite{NaTh19}, \cite{AbSa23}, and \cite{AbGh23} investigate \acs{TSN} and DetNet in the context of 5G networks. 
These networks have ultra-low latency requirements, e.g., to enable industrial automation.}
\subsubsection{Traffic Generation}
P4TG is a P4-based traffic generator for Ethernet/IP networks introduced by Lindner \textit{et al.}\cite{p4tg} that runs on an Intel Tofino™ switching ASIC. 
It is capable of generating both \acf{CBR} traffic and \FIi{Poisson} traffic at \SI{100}{\gbps} per port with more stable \FI{\acp{IAT}} than other traffic generators.
The implementation of Lindner \textit{et al.} does not suffer from the limitations of software-based traffic generators running on general-purpose CPUs, i.e., low data rates and \FI{being prone} to significant fluctuations in traffic generation\cite{pktgen} while being significantly less expensive than hardware-based commercial solutions.
The generated traffic can be fed back from the output ports to the input ports through other equipment.
\FI{In this process, P4TG} measures rates, frame types, frame sizes, and packet loss directly in the data plane, and samples RTTs and IATs in the control plane.
We leverage P4TG in \sect{eval} for \FIi{performance} evaluation.

%% file: chapters/03-psfp.tex
\section{Per-Stream Filtering and Policing (PSFP)}
\label{sec:psfp}
This section explains \acs{PSFP} as defined in IEEE Std 802.1Qci\cite{qci}.
First, an overview of \acs{PSFP} is given.
Then the \acs{PSFP} components, i.e., stream filter, stream gate, and flow meter, are explained.
\subsection{Overview}
\ac{PSFP} is a mechanism to ensure that streams of participants in a \acs{TSN} network adhere to the \FIi{\revieweri{stream} descriptors indicated during} admission control.
\FIi{Streams that do not adhere to the \revieweri{stream} descriptors \FIii{can be blocked entirely in a PSFP-enabled bridge or have individual violating frames dropped.}}
The \acs{PSFP} mechanism as defined in IEEE Std 802.1Qci\cite{qci} is \FIii{intended for} Ethernet bridges and consists of the three components stream filter, stream gate, and flow meter, that are processed in sequential order.
An overview is given in \fig{pdfs/psfp}. 

\figeps[0.4\textwidth]{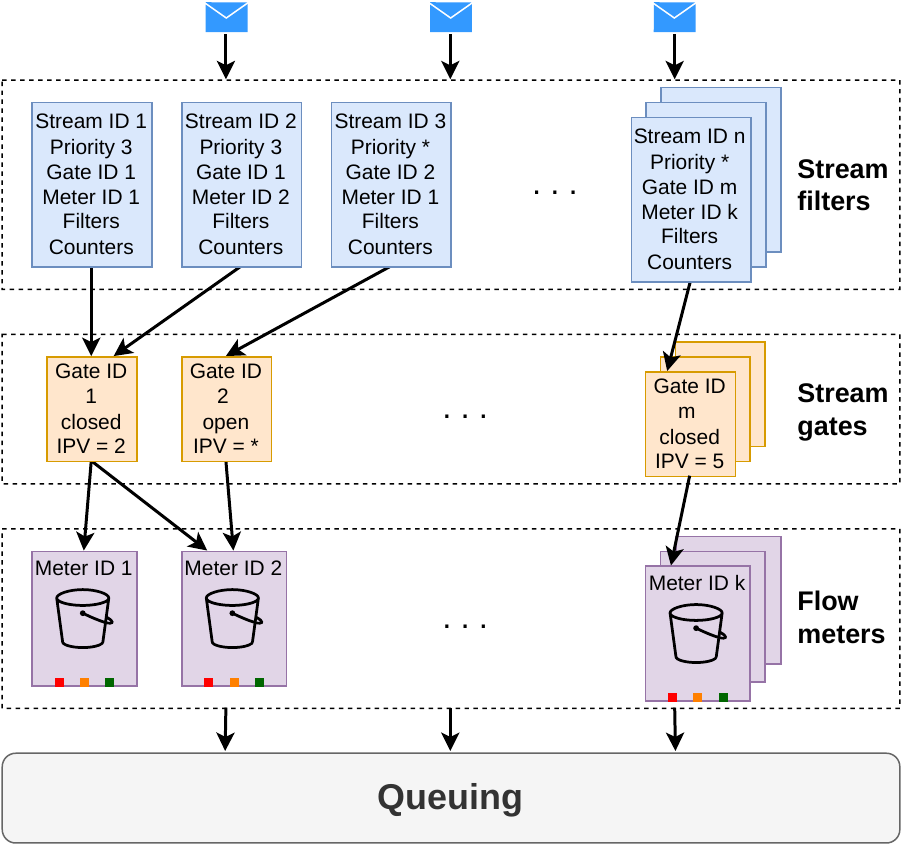}{The three components stream filter, stream gate, and flow meter compose the \acs{PSFP} mechanism. They are processed in sequential order\cite{qci}.}
 
An incoming frame is first identified by the stream filter instance.
The identified stream is then assigned \FIi{to} a stream gate instance and a flow meter instance. 
The stream gate decides whether to forward or drop the frame \FI{depending on its arrival time}, i.e., \FIi{whether} it adheres to a \FIi{cyclic} schedule.
Afterward\FI{s}, the flow meter checks if \FI{the frame} is in-profile in terms of bandwidth. 
 
\subsection{PSFP Components}
This section describes the purpose and functionality of each of the three \acs{PSFP} components \FI{and summarizes the conditions that must be met for a frame to be forwarded}.

\subsubsection{Stream Filter}
The stream filter instance implements several stream identification functions according to IEEE Std 802.1CB\cite{cb}.
\FIi{A stream identification function maps frames of a particular stream to a parameter called stream handle by identifying the stream according to header fields of a frame.
Stream identification functions in IEEE Std 802.1CB match, among others, the source MAC address and the VLAN ID, or IP header fields.}
A stream gate and a flow meter instance are assigned \FIi{to a frame} based on the stream handle.
\FI{If a frame is identified by a stream identification function, it is forwarded to the assigned stream gate instance.
Otherwise, \acs{PSFP} is not applied to the frame and it is queued with no further action.
For a frame that is not processed in PSFP, the queue is selected according to the frame priority in the \acs{PCP} field as defined in IEEE Std 802.1Qci\cite{qci}}.
Optional filtering functions can be defined, e.g., a maximum frame size filter that drops frames which exceed a certain frame size. 
In addition, the stream filter \FI{has an additional parameter called \textit{MaxSDUExceeded}} which can permanently block a stream if a frame exceeds \FI{the configured} maximum frame size once.
This prevents misbehaving participants from communicating in the \acs{TSN} network.
\subsubsection{Stream Gate}
\label{sec:stream_gate}
A stream gate monitors the \FI{arrival time} of frames and ensures that frames are only transmitted during their allowed time slices.

\FIi{A stream gate instance is controlled by a stream \acf{GCL}.
The stream \acs{GCL} in PSFP is similar to the \acs{GCL} of the TAS, but they are not the same.
Both, the stream \acs{GCL} and the TAS \acs{GCL} are composed of cyclically repeated time slices associated with a gate state that can be either open or closed.
In contrast to the TAS \acs{GCL}, the stream \acs{GCL} is independent of the egress port and its priority queues.
It does not contain the 8-bit vector to control queue-specific gates.
Instead, the stream \acs{GCL} only \FIii{encodes the time slices} of a single stream gate instance which corresponds to one or more identified streams.
Frames arriving \FIii{at the switch} during a time slice in an open state of the stream \acs{GCL} are forwarded, \FIii{frames arriving in} the closed state are dropped.
In contrast to the TAS, the PSFP stream gate cannot delay frames, it can only drop \FIii{them}.
Additionally, time slices in a stream \acs{GCL} have an optional value called \ac{IPV}. 
This value is used to select the egress queue for a frame in this bridge.
If \FI{the \acs{IPV}} is not set, the \ac{PCP} value from the 802.1Q header is used.}

\input{chapters/psfp_params_table}

\FI{Stream \acp{GCL} must be carefully planned in advance.
They \FIi{are} periodic to \FIi{facilitate} continuous operation.}
The period $p_i$ of a stream \acs{GCL} $g_i$ is the duration of all individual time slices in the stream \acs{GCL} $g_i$.
Multiple stream \acp{GCL} are usually aggregated into a hyperperiod $h$ which equals the \ac{LCM} of all periods. 
This is needed to model the cyclic behavior\cite{PoRa17, RaPo17, GaVo18} and is illustrated, \FIi{together} with a stream \acs{GCL}, in \fig{pdfs/hyperperiod}.

\figeps[\columnwidth]{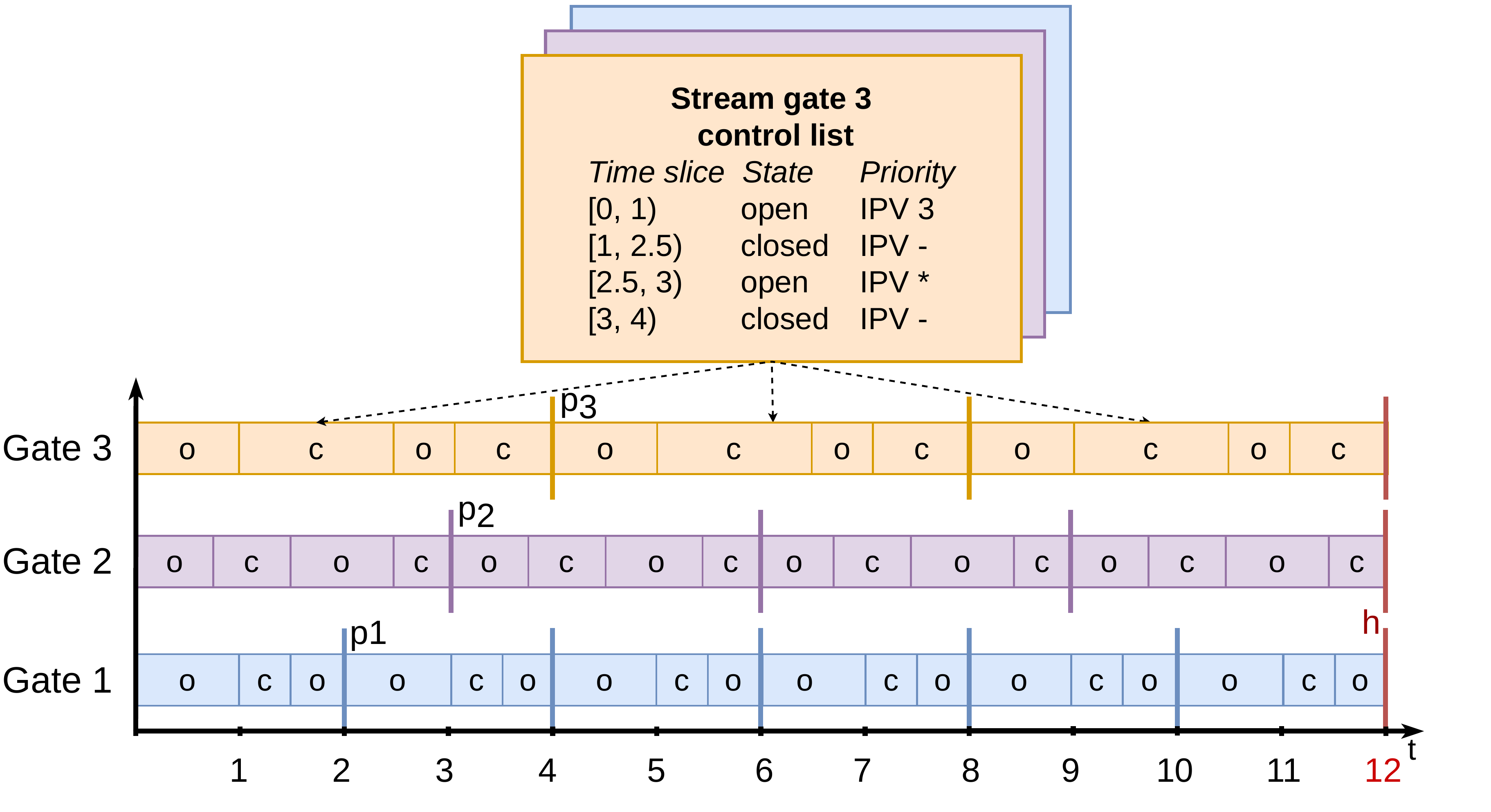}{Example hyperperiod consisting of three stream \acp{GCL} with periods $p_1 = 2$, $p_2 = 3$, and $p_3 = 4$, and time slices in the open (o) or closed (c) state. Each stream \acs{GCL} is extended to the least common multiple of $h=12$ to form a hyperperiod.}

The shown hyperperiod contains three stream gate instances, each with its own stream \acs{GCL} and its own period.
Each stream \acs{GCL} is extended to the \acs{LCM} of 12 time steps.

Furthermore, a stream gate \FI{holds additional parameters to} permanently \FI{close the gate} if it receives a single frame during a time slice in a closed state \FI{(\textit{GateClosedDueToInvalidRX})}, or if it receives too much data in a single time slice \FI{(\textit{GateClosedDueToOctetsExceeded})}.
\subsubsection{Flow Meter}
\label{sec:flow_meter}
The flow meter instance monitors the compliance of \acs{TSN} talkers with their admitted \FIi{rate in their \revieweri{stream} descriptor.}
A two-rate, three-color marking token bucket policer conforming to RFC 2698\cite{rfc2698} is used for this purpose.
\revieweri{The token bucket policer is illustrated in \fig{pdfs/token_bucket}.}

\figeps[0.4\textwidth]{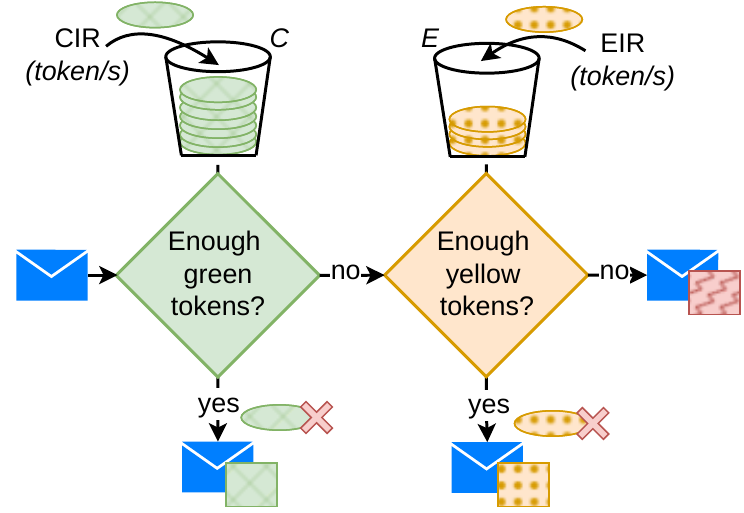}{A two-rate, three-color marking token bucket policer according to RFC 2698\cite{rfc2698}.}

\revieweri{In the two-rate three-color marking token bucket algorithm, two buckets $C$ and $E$ are filled with tokens at the \ac{CIR} and \ac{EIR} respectively.
The two buckets have a capacity of Committed Burst Size (CBS) tokens, and Excess Burst Size (EBS) tokens.
The \ac{CIR} corresponds to the guaranteed bandwidth, and the \ac{EIR} represents the additional bandwidth that can be used temporarily if available.
A token bucket is initially completely filled.
An incoming frame consumes tokens from the buckets according to its frame size.
If there are enough tokens in bucket $C$, the frame is marked green and the tokens are removed from the bucket.
If there are not enough tokens in bucket $C$ but there are enough in bucket $E$, the frame is colored yellow and consumes tokens from bucket $E$.
Otherwise, the frame is marked red.
In RFC 2698, the color of a frame is reflected in the DS field of the IP header. 
However, this concept can also be used for non-IP traffic by using other header fields to reflect the color.}

The RFC only describes how to color frames, but does not associate a color with an action.
In \acs{PSFP}, green-colored frames are forwarded.
For yellow frames, the \ac{DEI} flag\footnote{The \acs{DEI} flag indicates frames that may be dropped in the presence of congestion.} is set in the 802.1Q header before they are forwarded, and red frames are dropped\cite{qci}.

Additional parameters are added to the flow meter instance, e.g., the \textit{DropOnYellow} parameter, which additionally drops yellow frames, or the \textit{MarkAllFramesRed} parameter, which permanently blocks the flow meter once the assured bandwidth has been exceeded. The \textit{ColorMode} parameter indicates whether or not a frame pre-color should be respected.
\subsubsection{Summary of PSFP Parameters}
\label{sec:psfp_summary}

\FI{
The optional parameters available for PSFP components are summarized in \tabl{psfp_params}.
A frame that is identified by a stream identification function \FIi{is assigned to a stream gate instance and a flow meter instance. 
The frame} must meet all of the following conditions to be forwarded.}

\FI{
\begin{enumerate}
    \item The frame size is below the maximum frame size.
    \item The identified stream is not permanently blocked due to a previous frame exceeding the maximum frame size if the MaxSDUExceeded parameter is enabled.
    \item The frame is assigned to a time slice of the stream \acs{GCL} in an open state.
    \item The assigned stream gate is not permanently closed due to a previous frame if the GateClosedDueToInvalidRX parameter is enabled.
    \item The number of bytes transmitted in the current time slice of the stream \acs{GCL} does not exceed the maximum configured number of bytes per time slice if the GateClosedDueToOctectsExceeded parameter is enabled.
    \item The frame is not marked red by the flow meter.
    \item The assigned flow meter is not permanently blocked due to a previous frame if the MarkAllFramesRed parameter is enabled.
    \item The frame is neither pre-colored yellow\footnote{A frame is pre-colored yellow if it has the \ac{DEI} flag already set upon entering the switch.} nor marked yellow by the flow meter if the DropOnYellow parameter is enabled.
\end{enumerate}}

%% file: chapters/psfp_params_table.tex
\begin{table*}[htb!]
	\caption{Summary of available optional parameters for PSFP components in IEEE Std 802.1Qci\cite{qci}. \FIi{They are all} implemented in P4-PSFP.}
	\begin{center}\resizebox{1\textwidth}{!}{
 \FI{
\begin{tabular}{|l||l|l|c|}
\hline
\textbf{Parameter name}                & \textbf{PSFP component} & \textbf{Description}                                                                                                                          & \multicolumn{1}{l|}{\textbf{\begin{tabular}[c]{@{}l@{}}Implemented\\  in P4-PSFP\end{tabular}}} \\ \hline \hline
\textit{MaxSDUExceeded}                & Stream filter           & \begin{tabular}[c]{@{}l@{}}The stream is permanently blocked if a single frame \\ exceeds the configured frame size.\end{tabular}             & \checkmark                                                                       \\ \hline
\textit{GateClosedDueToInvalidRX}      & Stream gate             & \begin{tabular}[c]{@{}l@{}}The gate is permanently closed if a single frame\\ arrives during a time slice in a closed state.\end{tabular}      & \checkmark                                                                       \\ \hline
\textit{GateClosedDueToOctetsExceeded} & Stream gate             & \begin{tabular}[c]{@{}l@{}}The gate is permanently closed if too many bytes\\ are sent during a single time slice.\end{tabular}                 & \checkmark                                                                       \\ \hline
\textit{DropOnYellow}                  & Flow meter              & \begin{tabular}[c]{@{}l@{}}Yellow-labeled frames, either pre-colored or colored\\ by the flow meter, are marked red and dropped.\end{tabular} & \checkmark                                                                       \\ \hline
\textit{MarkAllFramesRed}              & Flow meter              & \begin{tabular}[c]{@{}l@{}}The flow meter is permanently blocked if a single frame\\ is marked red.\end{tabular}                              & \checkmark                                                                       \\ \hline
\textit{ColorMode}                     & Flow meter              & \begin{tabular}[c]{@{}l@{}}The color mode indicates whether the pre-color of a\\ frame should be respected.\end{tabular}                             & \checkmark                                                                       \\ \hline
\end{tabular}}}
	\label{tab:psfp_params}
	\end{center}
\end{table*}

%% file: chapters/04-p4.tex
\section{Introduction to P4}
\label{sec:p4}
This section provides fundamentals of the P4 programming language. 
We start with an overview of P4. 
Then we explain \acp{MAT}.
Finally, we explain the P4 pipeline in \FIi{the} context \FIi{of} the \ac{TNA}.
\subsection{Overview}
\ac{P4} is a high-level programming language for describing a data plane and an architecture-based programming model.
The P4 language can be used to manipulate packets and make forwarding decisions in the data plane to implement user-defined algorithms. 
Devices that implement a specific architecture are called targets. 
Targets can be either software-based, e.g., the \texttt{simple\_switch} as used in the BMv2 \cite{bmv2}, or hardware-based, e.g., the Intel Tofino™ switching ASIC\cite{tofino}.
A P4 program is designed for a specific architecture and may be used with all targets that implement this specific architecture\cite{p4}.
Externs extend the functionality of a P4 program with target-specific operations such as registers or counters. 
The actual implementation of the extern functionality depends on the target.
One hardware-based architecture is the \ac{TNA} which will be discussed further in this section. 
It builds upon the more basic \ac{PSA}\cite{psa}. 

\FI{Implementing a complex, time-dependent algorithm on real hardware poses many challenges.
To meet the line rate of the switch, hardware targets often have limitations on the number of operations that can be applied per packet, so writing a P4 program for them is more difficult than for software targets.
Implementations on software switches or simulation frameworks, like in CoRE4INET\cite{core4inet}, do not suffer from these constraints and \FIi{cannot guarantee line rate.
Therefore they are not suitable for} a real-world setting.}
Further information on P4 can be found in a detailed survey by Hauser et al.\cite{kn}.
\subsection{Match+Action Tables (MATs)}
A core feature of P4 programs are match+action tables (MATs). 
They describe tables that associate key fields, i.e., header or metadata fields, with user-defined actions.
\FIi{Metadata in P4 can be intrinsic \FIii{or user-defined.
Intrinsic metadata contains data provided by the architecture, e.g., the ingress port of a frame}.}
The concept of \acp{MAT} is illustrated in \fig{pdfs/mat}.

\figeps[\columnwidth]{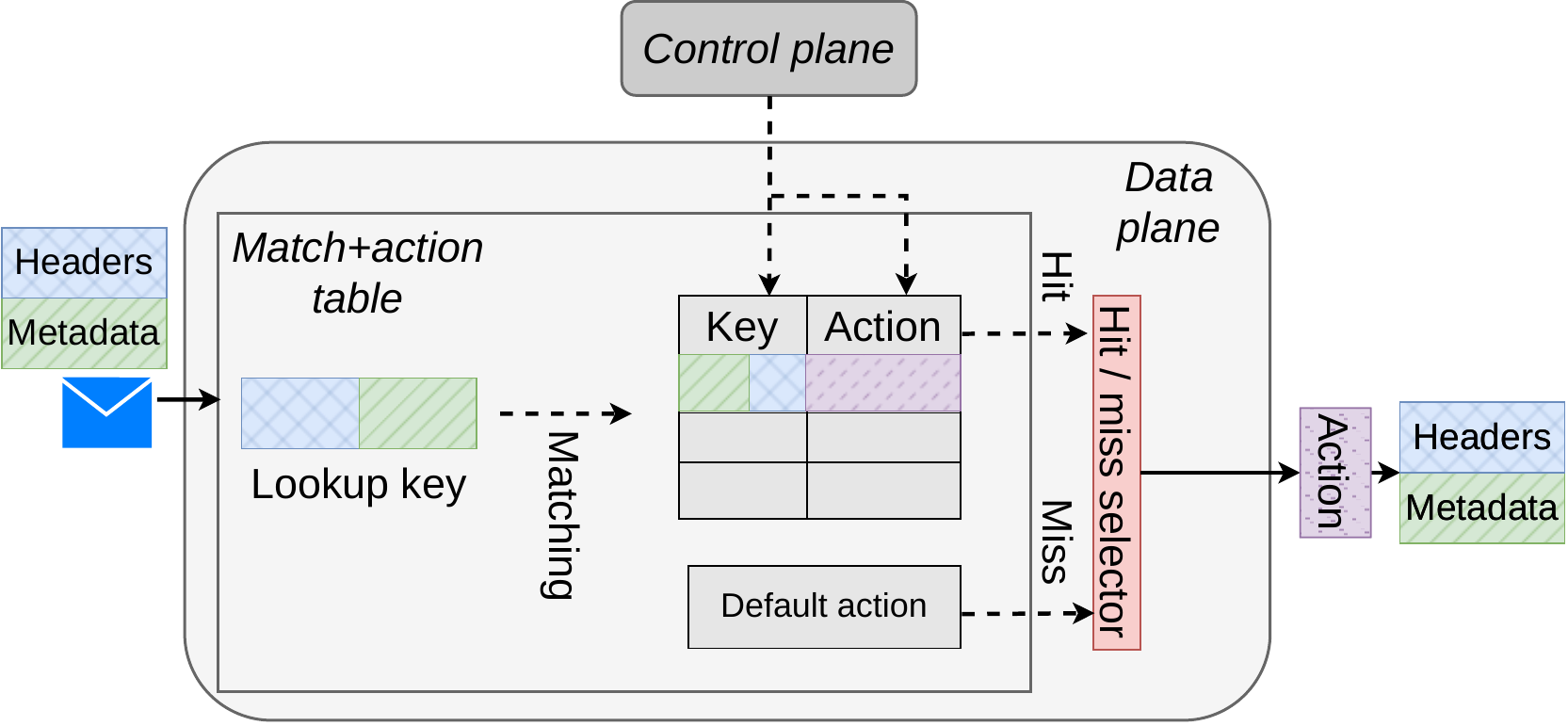}{\FI{Illustrated \acf{MAT}\cite{kn}}. The header and metadata fields of a frame form a composite key to match entries in the table. In the data plane, either the assigned action is executed on a match, or the default action is executed on a miss. The table entries are populated by the control plane.}

An incoming packet is matched against the key fields in a \acs{MAT} and an associated action is executed upon matching an entry. 
Actions can be described as small code fragments \FIi{and} may affect packet forwarding or manipulate fields. 
\acp{MAT} and actions are defined in the data plane while the entries of a \acs{MAT} and the parameters of an action must be filled in by the control plane.
The core P4 language defines three match types for MATs: ternary, \ac{LPM}, and exact. 
A ternary match applies a predefined bitmask to the key fields.
\FI{An \acs{LPM} match type implements longest prefix matching which is well known from IPv4.}
The exact match type can be seen as a special case of \acs{LPM} with the maximum prefix length.
\subsection{Tofino Native Architecture (TNA)}
In general a \acs{P4} pipeline consists of a programmable packet parser and control blocks, followed by a packet deparser. 
The packet parser extracts header information from packets and stores the information in an adequate data structure. 
Control blocks define packet processing operations and describe the algorithm to be applied to packets \cite{p4spec}.
P4 control blocks may use branching constructs as well as logical and simple arithmetic expressions.
Loop constructs cannot be used in P4.
Recirculation is a mechanism that models loop behavior by sending a packet from the output port back to the input port where it traverses the pipeline again. 

In particular, the \FI{pipeline of the} \acs{TNA}, the architecture used by the Intel Tofino™, consists of both ingress control blocks and egress control blocks\cite{tna}, each of which has its own parser and deparser. 
\FI{The TNA pipeline} is illustrated in \fig{pdfs/tna}. 

\figeps[\columnwidth]{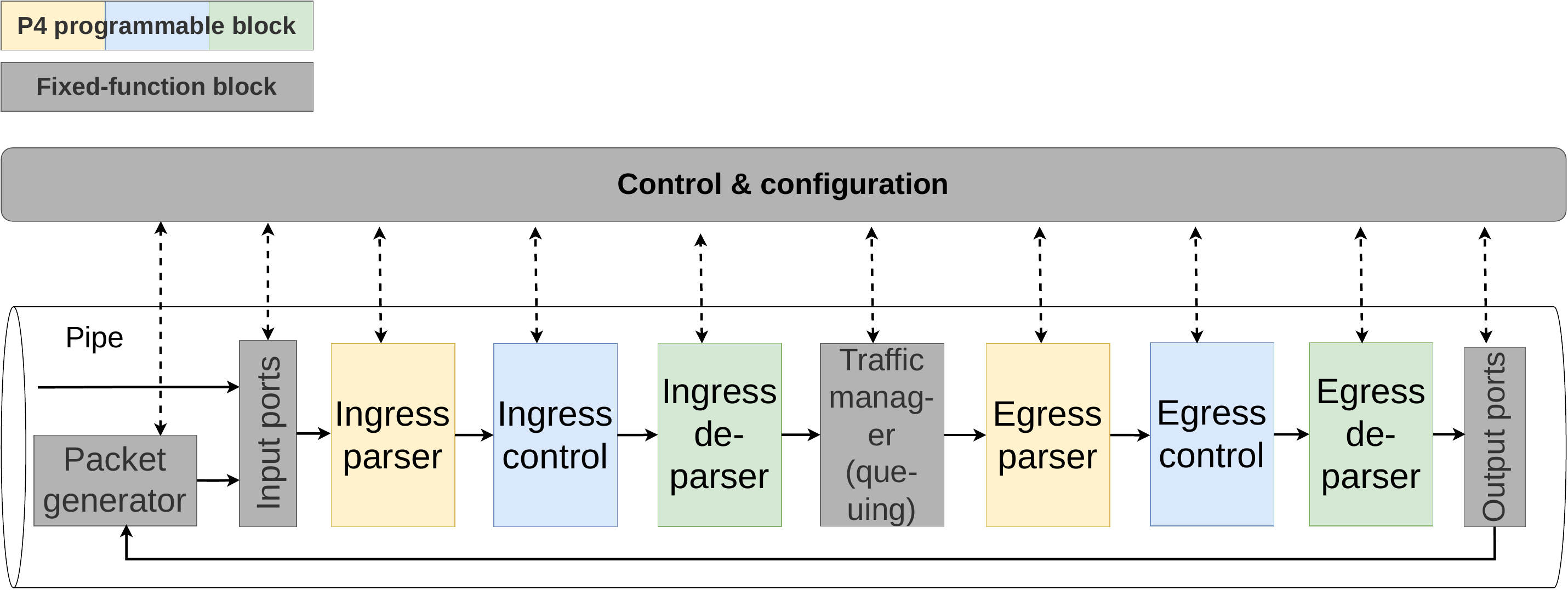}{\FI{Illustrated TNA pipeline\cite{tna}}. It consists of several P4 programmable blocks in the ingress and egress and some fixed-function blocks, such as the packet generator.}

 \begin{figure*}[htb!]
  \begin{center}
   \leavevmode
      \parbox[t]{\textwidth}{%
        \resizebox{\textwidth}{!}{\includegraphics{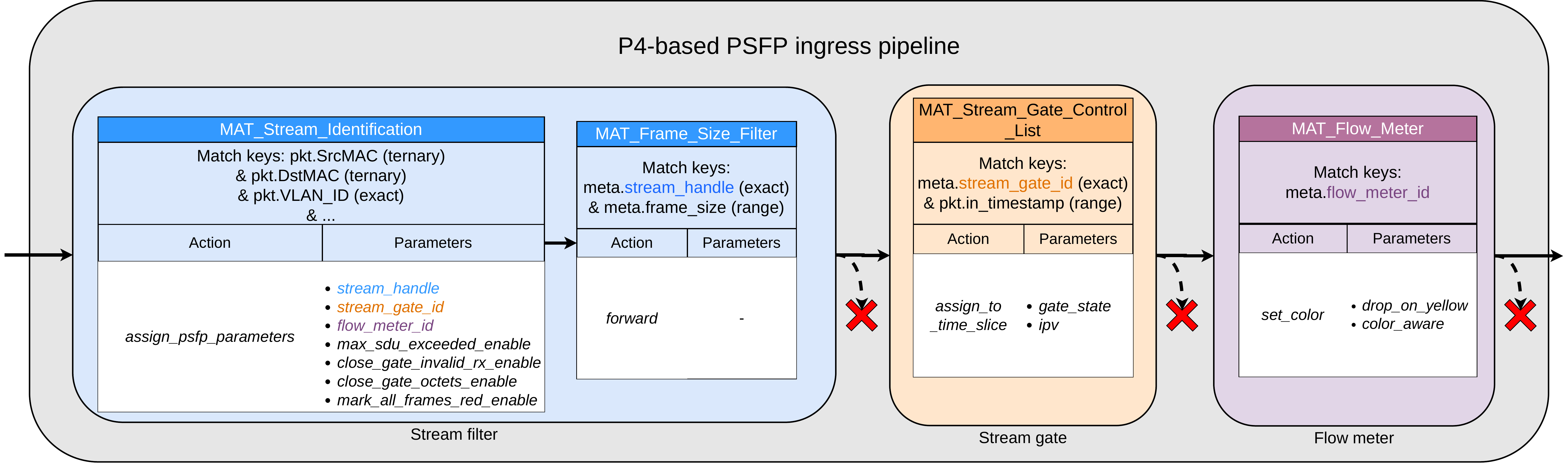}}
      }
   \caption{\FI{Overview of the P4-based PSFP ingress pipeline.} A frame is identified by the stream filter via a MAT, filtered according to its frame size, assigned to a stream gate time slice by its ingress timestamp, and finally flow metered. Any of the components may drop the frame.\vspace{-0.2cm}}
   \label{fig:implementation_overview}
  \end{center}
 \end{figure*}

Parsers, deparsers, and both control blocks are programmable in P4 and can be used to control packet forwarding and manipulation.
A \acs{TNA}-specific packet generator can be configured to assemble packets and send them through the Intel Tofino™ output ports, e.g., in a periodic manner.
Furthermore, the \acs{TNA} defines an additional type for matching in \acp{MAT}: range. 
This type can be used to match a key field to an interval range and execute the action accordingly.

%% file: chapters/05-implementation.tex
\section{P4 Implementation of PSFP for Intel Tofino™}
\label{sec:implementation}
In this section, we give an overview of the implementation of \acs{PSFP} in P4 on an Intel Tofino™ switching ASIC integrated within an Edgecore Wedge 100BF-32X\cite{tofino} switch. 
P4 language-specific challenges such as the periodicity of stream \acp{GCL} are addressed and solutions are presented. 
An approach to account for time inaccuracy in implementations without time synchronization capabilities in the data plane is proposed.
The source code is publicly available on GitHub\cite{git}.
\subsection{Overview}
\label{sec:implementation_overview}
A high-level description of the P4-PSFP implementation is given in \fig{implementation_overview}.
\revieweri{A frame entering the P4-PSFP bridge is matched against a \acs{MAT} in the stream filter component.
For simplicity, we only consider VLAN-tagged frames in our implementation.
The key in the stream filter \acs{MAT} consists of selected header fields to identify the stream, e.g., the VLAN ID.
The header fields used for this purpose depend on the used stream identification function.
If the frame is identified by the stream filter, it belongs to a time-sensitive stream and is eligible for \acs{PSFP} processing.}
As a result, the stream gate instance and flow meter instance are assigned to the frame \FIi{using metadata}.
Furthermore, the frame size is checked in an additional \acs{MAT}.

In the next step, the frame is matched \FI{depending on the frame ingress timestamp} against a \acs{MAT} in the stream gate instance.
\FI{This MAT represents the stream \acs{GCL} and a frame is assigned to a time slice}.
\FI{The associated action \FIii{retrieves the gate state} in the current time slice and assigns the \acs{IPV} to the frame.}
\FI{Each entry in the stream gate instance MAT corresponds to a time slice in the open state of a stream \acs{GCL}. 
\revieweri{Time slices in the closed state do not need to be listed in the stream gate instance \acs{MAT}.
A gap between two open time slices in the stream gate instance \acs{MAT} implicitly models closed time slices.
Packets arriving in such a gap do not match in the \acs{MAT}, resulting in a miss.
On a miss in the \acs{MAT}, the packet is dropped by the default action.}}
If the frame was assigned to a time slice in an open state, it is further matched against a \acs{MAT} with its previously assigned flow meter ID in the flow meter instance.

In the flow meter instance, a token bucket algorithm according to RFC2698\cite{rfc2698} is applied to the frame.
\FI{To that end}, a P4-specific meter extern is assigned to the flow meter \acs{MAT}.
\FI{This meter extern} can be configured with the \FI{appropriate} rates for the token bucket algorithm.
Depending on the frame color, the frame is then dropped, marked, or forwarded.
Any of the components may drop a frame if it violates conditions announced during admission control, i.e., it exceeds a certain frame size, arrives in a closed state time slice, or exceeds the admitted bandwidth.

\subsection{Details of the PSFP Implementation}
This section gives insights into the details of \FI{the} implementation. 
First, some challenges and their solutions are elaborated. 
Subsequently, the interaction of all implemented mechanisms is explained.
\subsubsection{Maximum Frame Size Filter}
\label{sec:max_filter}
\FI{The stream filter component is designed to optionally discard frames that exceed a configured maximum frame size. 
The \acs{TNA} provides the frame size in the intrinsic egress metadata for further processing.
Since \acs{PSFP} has the ability to modify frame priorities and thus their queue, the \acs{PSFP} mechanism must be applied prior to the queuing process.
In the \acs{TNA}, queuing takes place subsequent to the ingress control block but preceding the egress control block, as illustrated in \fig{pdfs/tna}.
As a result, the implementation of \acs{PSFP} must reside within the ingress control block.
\FI{However,} the frame size is only available after egress parsing in the \acs{TNA}\cite{tna}, but is needed for filtering in the ingress control block in the stream filter.}

We use the concept of recirculation \FI{in P4} to solve this problem.
A frame that is eligible for \acs{PSFP} processing traverses the P4 pipeline twice. 
On the first pass, the frame gets a recirculation header prepended to its header stack in the egress control block.
This header contains the frame size available in the egress control block.
The frame is then recirculated and sent back to an ingress port.
After recirculation, the frame size in the recirculation header can be used to apply the stream filter in the ingress control block.
Packet processing in the Intel Tofino™ takes a constant amount of time \reviewerii{on sub-microsecond scale}, so a recirculation only adds a constant, known amount of time to the processing delay.
\reviewerii{The effect of recirculation in P4-PSFP is discussed in \sect{discuss_recirculation}.}
\subsubsection{Periodicity of Stream GCLs}
\label{sec:periodicity}
This section focuses on the implementation of \FIi{configured} \FIi{stream \acp{GCL}} and achieving periodicity. 
Finding a valid schedule and scheduling algorithms for a TSN environment are outside the scope of this work.
Further information on scheduling algorithms in TSN networks can be found in a detailed survey by Stüber \textit{et al.}\cite{StOs23}.

A \acs{PSFP} \FIi{stream \acs{GCL}} is defined by \FI{\FIi{time slices} with relative time steps} that are periodically repeated.
The \FIi{stream \acs{GCL}} must be able to operate continuously for an indefinite amount of time.
To that end, the absolute timestamp of an incoming frame must be mapped to the correct relative time slice in the \FIi{stream \acs{GCL}}. 
This is illustrated in \fig{pdfs/hyperperiod_2}.

\figeps[\columnwidth]{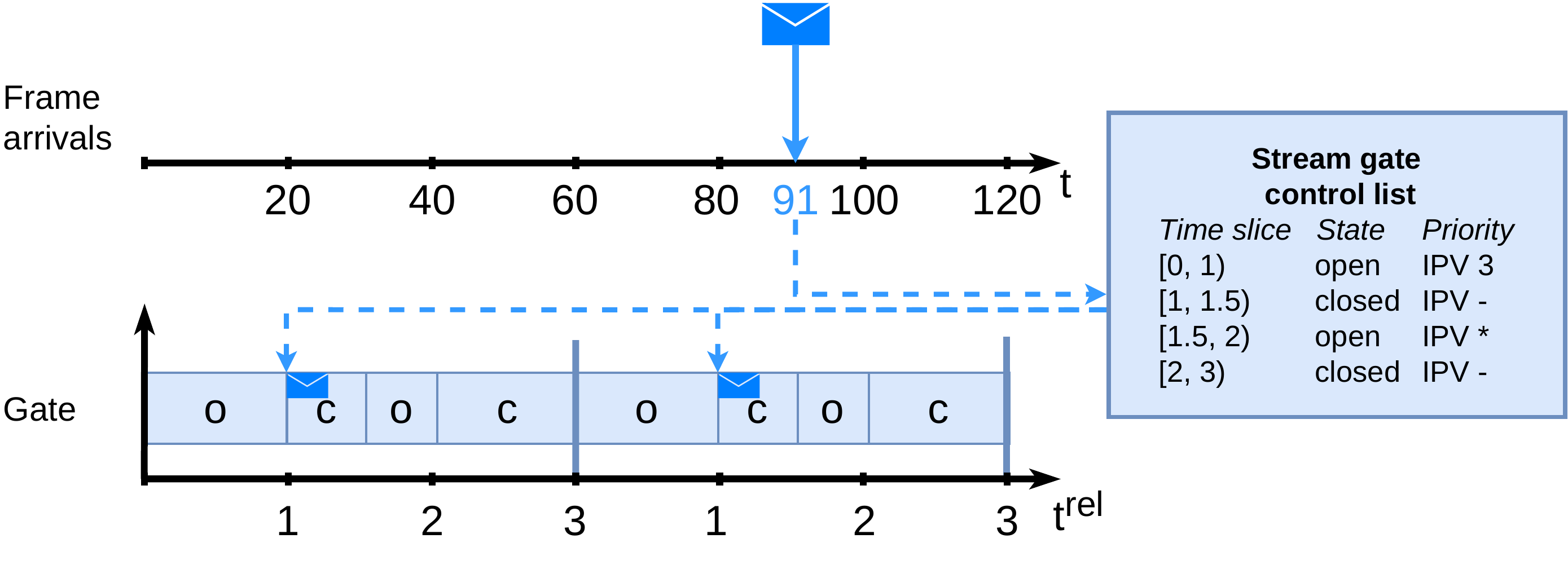}{A frame received at an arbitrary absolute timestamp, e.g., at $t_i = 91$, \FI{is} mapped to its relative position in the \FIi{stream \acs{GCL}}, e.g., to the time slice $[1, 1.5)$.}

\FIi{Multiple stream \acp{GCL} are aggregated to a hyperperiod to model cyclic behavior as described in \sect{stream_gate}}.
\FI{The $i$-th frame is received at time $t_i$.
For further processing, i.e., matching a frame to a time slice in the stream gate instance, a mapping to a relative frame arrival time $t^{rel}_i$ in a hyperperiod with the duration $h$ is required.
Let $t_0^h$ be the start time of the first hyperperiod.
Then its relative arrival time can be determined as shown in \equa{rel_modulo}.
\begin{align}
    t^{rel}_i = (t_i-t_0^h) \text{ mod } h \label{eq:rel_modulo}
\end{align}
The calculation in \equa{rel_modulo} cannot be implemented on the Intel Tofino™ due to the lack of a modulo operator.}

We leverage the \FI{internal} packet generator of the \acs{TNA} to \FI{imitate} the modulo operator.
The packet generator is configured to periodically generate a \FI{packet \FIi{$j$} every $h$ time steps and send it to an ingress port of the P4-PSFP switch.}
On reception of such a generated packet, its ingress timestamp \FI{$t_j^h$} is stored in a data plane register \FI{$r^h$} that references the last completed hyperperiod.
The relative position $t^{rel}_i$ of a frame $i$ within a hyperperiod of the duration $h$ can then be calculated \FI{as described in \equa{t_rel}}.
\begin{align}
    t^{rel}_i &= t_i - t_j^h \label{eq:t_rel}
\end{align}
\reviewerii{\FI{\equa{t_rel}} is semantically equivalent to \FIi{the} modulo operation \FIi{in \equa{rel_modulo}} with the requirement that the timestamp of the generated hyperperiod packet arrives perfectly after an elapsed hyperperiod, i.e., it holds that $t_j^h = t_1^h + (j-1)\cdot h$.
Since this may not always be the case, we introduce a mechanism to account for such inaccuracies in \sect{time_sync}.}

The Intel Tofino™ uses integers to represent timestamps with a granularity of \SI{1}{\nano\second}\cite{tna}.
Moreover, timestamps on the Intel Tofino™ are limited to 48 bits.
Thus, they overflow after \reviewerii{$t^{max} = 2^{48}-1 \text{ ns} \approx 3.25$ days}.
This leads to an underflow in the calculation of the relative frame arrival time.
\reviewerii{\fig{pdfs/overflow} illustrates the problem of underflows in the calculation of the relative timestamp if $t^{max}$ is reached.}

\figeps[\columnwidth]{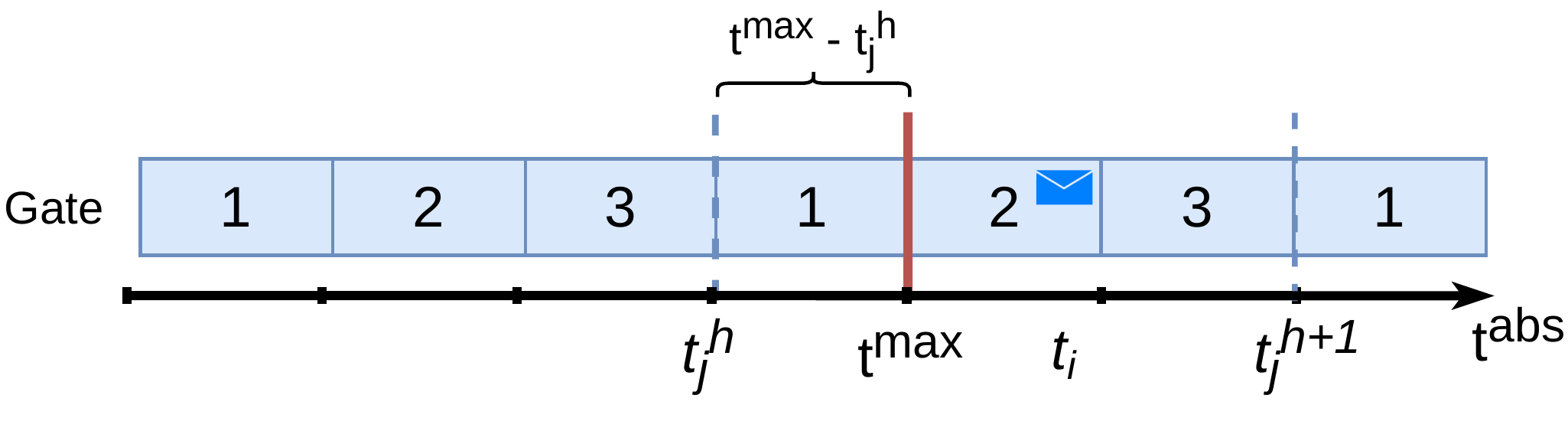}{\reviewerii{The relative timestamp $t_i^{rel}$ of a frame $i$ arriving at $t_i$ after $t^{max}$ was reached results in an underflow if the timestamp of the last elapsed hyperperiod $t_j^{h}$ was set before $t^{max}$.}}

\reviewerii{
In \fig{pdfs/overflow}, the absolute arrival time $t_i$ of frame $i$ is smaller than the stored timestamp of the last completed hyperperiod $t_j^h$ because $t$ overflowed at $t^{max}$.
Therefore, the calculated relative position $t_i^{rel}$ results in a negative value according to \equa{t_rel}, i.e., it underflows.
As the data type of the timestamp is unsigned, the negative value is represented as a very large value.
An underflow is therefore detected if the calculated relative timestamp $t_i^{rel}$ results in a very large value, i.e, larger than the hyperperiod $h$.
If an underflow is detected, the elapsed time of the hyperperiod before $t^{max}$ was reached is calculated and added to the arrival time $t_i$ of the frame as shown in \equa{overflow}.}
\begin{align}
    t^{rel}_i &= t^{max} - t_j^{h} + t_i \label{eq:overflow}
\end{align}

The packet generator of the TNA can be configured with eight different \FI{periodic} triggers\cite{tna}, i.e., eight different \FIi{hyperperiods} can be implemented \FI{on a P4-PSFP switch}.
\FIii{Each hyperperiod comprises multiple stream gate instances with individual stream \acp{GCL} that accommodate multiple streams.}
\FIi{The packet generator configures the periodic trigger for a hyperperiod on a per-ingress-port basis. 
Therefore, P4-PSFP supports eight ingress ports with distinct hyperperiods.
On each ingress port, the individual stream \acp{GCL} of the stream gate instances yield the hyperperiod of this ingress port.}
\label{sec:hyperperiod}
\subsubsection{Timestamp Matching Precision}
The assignment of a frame to a time slice in a \FIi{stream \acs{GCL}} is based on its ingress timestamp.
The implementation selects \FIi{20 bits} from the \SI{48}{\bit} wide timestamp of an incoming frame, \FI{e.g., bit 12 to 31}, to enable matching in the stream gate \acs{MAT} using the range match type.
The \acs{MAT} timestamp key is truncated to these \FI{selected} bits.
\FIi{This is necessary to save computing resources in a hardware-constrained switch.}
The truncation process reduces the potential temporal resolution of \FIi{time slices in a \FIi{stream \acs{GCL}}}.
\FI{P4-PSFP} supports a temporal resolution of \SI{2}{\micro\second} to \FI{about} \SI{2.1}{\second}, i.e., a time slice can be assigned down to a precision of \SI{2}{\micro\second} and a hyperperiod can be at most \SI{2.1}{\second} long. 
Although this is technically a limitation of the implementation, it is not a problem.
According to a survey on scheduling algorithms for the \acs{TAS} by Stüber \textit{et al.} \cite{StOs23}, stream hyperperiods range from \SI{32}{\micro\second} in \cite{XiXi20} to \SI{500}{\milli\second} in \cite{CrOl16}.
\FI{The range of selected bits from the timestamp can be adjusted to the actual requirements of the network environment to achieve nanosecond accuracy.}
\subsubsection{\FI{Permanent Stream Blocking}}
Each PSFP component must be able to permanently block traversing streams that \FIi{do not conform to the \FIi{\revieweri{stream} descriptors indicated during} admission control.}
The stream filter can block a stream if the stream exceeds the maximum frame size.
The stream gate can be closed to block streams if frames are transmitted during a time slice in a closed state \FI{or if too many bytes arrive in a single time slice.}
The flow meter can be blocked if \FIi{the streams using it} exceed the announced bandwidth.

Permanently blocking a stream in the stream filter, the closing of a stream gate, and the blocking of a flow meter are all handled entirely in the data plane. 
\FIi{As a result, there is no delay between a stream not conforming to its \revieweri{stream} descriptor and the P4-PSFP switch blocking the corresponding PSFP component.}
\FIii{The data plane keeps track of the blocking of each stream, each stream gate, and each flow meter in multiple registers.}

\FI{Additionally, to block the stream gate if too many bytes are transmitted in a single time slice, \FIi{the remaining number} of available bytes per time slice per hyperperiod is maintained in a separate register.
\FIi{This register is decremented by the frame size for each frame received and reset upon receiving a generated hyperperiod frame.}}
Frames are dropped according to the register states.
\FIi{The control plane configures which of the permanent stream blocking mechanisms is enabled.}
\subsubsection{Interaction of the Implemented Mechanisms}
\label{sec:p4_psfp_details}
There are four different paths a frame can pass in the data plane processing pipeline, combining all previously explained mechanisms.
The paths differ for different frame header stacks. 
They are illustrated in \fig{pdfs/implementation}. 

\figeps[\columnwidth]{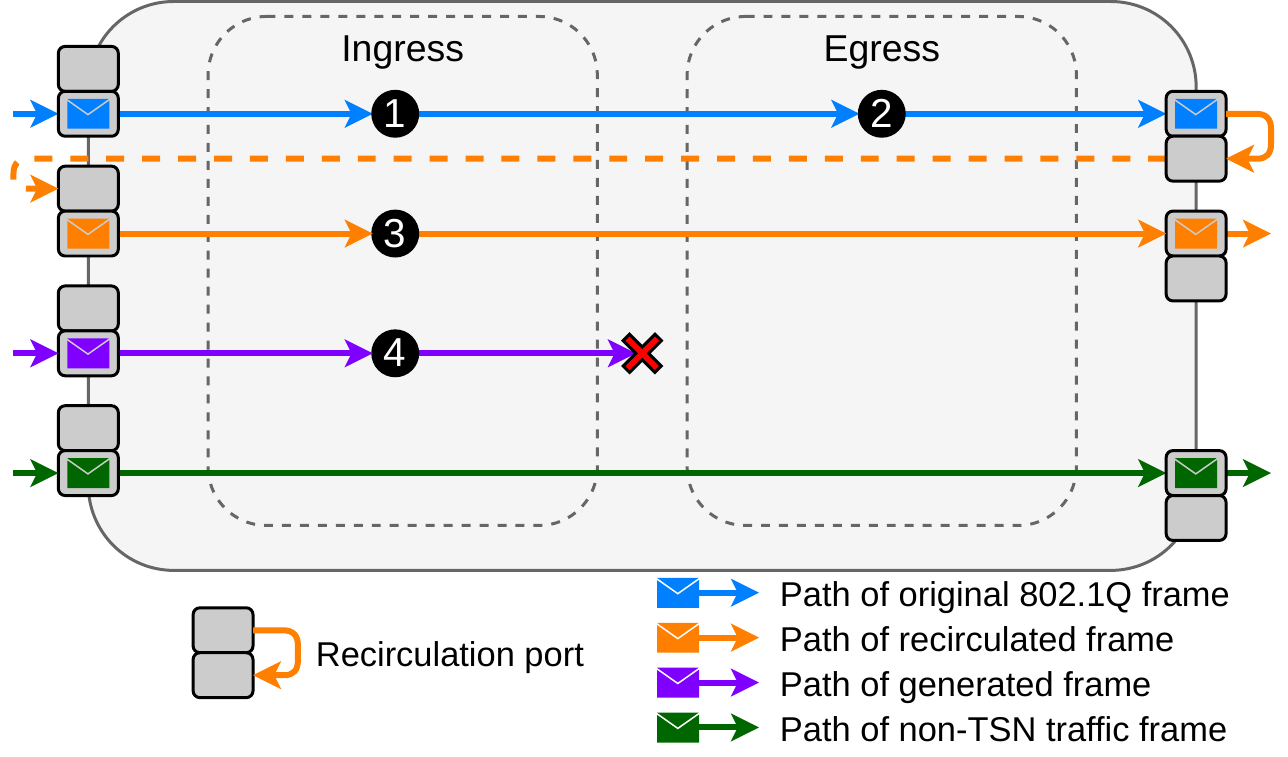}{A frame containing a VLAN tag first takes the blue path and traverses the pipeline once. It is then recirculated and PSFP is applied on the orange path before the frame is forwarded. Generated packets are dropped after ingress processing in the purple path. Non-TSN traffic is forwarded without any further action in the green path.}

A frame that has a VLAN tag is eligible for \acs{PSFP} processing. 
For such frames, a timestamp of the last completed hyperperiod is retrieved from a register in step \ballnumber{1}.
The relative position in the \FIi{stream \acs{GCL}} is calculated from this value and appended to the frame.
The frame is then forwarded to the egress control block.

In egress processing in step \ballnumber{2}, the frame size is added to a recirculation header, together with the relative position in the \FIi{stream \acs{GCL}}.
This increases the size of each frame by additional seven bytes.
Furthermore, the detection of overflows when calculating the relative position in a hyperperiod is handled.
The frame is recirculated and sent back to an ingress port.
The recirculation header can then be used to perform the \acs{PSFP} mechanism in the ingress control block after recirculation.

Finally, in step \ballnumber{3}, the PSFP mechanism is executed and the stream filter, stream gate, and flow meter instances are applied.
Each component is represented as a separate control block consisting of multiple MATs to load configuration parameters and apply the appropriate actions.
A frame needs to pass all of the conditions mentioned in \sect{psfp_summary} to be forwarded.

Frames in the purple path in \fig{pdfs/implementation} correspond to packets generated by the \acs{TNA} packet generator.
They are used to store a hyperperiod timestamp in a register $r^h$ in step \ballnumber{4} to account for the periodicity of \FIi{stream \acp{GCL}}.
After that, they are no longer needed and are therefore discarded.

Non-TSN traffic, i.e., frames that do not have a VLAN header, or are not identified by a stream filter entry, are treated as best-effort traffic and are forwarded without any further action.
\subsection{Time Synchronization}
\label{sec:time_sync}
High-precision time synchronization is essential to correctly assign a frame to its time slice in the \FIi{stream \acs{GCL}}. 
Although the Intel Tofino™ is capable of synchronizing to the network time\cite{tofino} with the \acf{PTP}\cite{ptp}, not all hardware platforms are capable of doing so.
One alternative to \acs{PTP} is the implementation of the \ac{DPTP}\cite{dptp}. 
However, this \FI{requires} additional space in the P4 pipeline and is not feasible due to the complexity of the P4-PSFP implementation.
This section first describes the problems of lacking time synchronization and then introduces a proposal for achieving time synchronization of \FIi{stream \acp{GCL}} in the data plane where the application of protocols like \acs{PTP} is not possible.
\subsubsection{Problem \FI{S}tatement}
\label{sec:clock_dift_problem}

\FIi{The problems resulting from a lack of time synchronization are manifold. 
We identify and present three root causes and their implications on the time synchronity of stream \acp{GCL}.}

\FIi{First,} the data plane \FIi{of P4-PSFP and its control plane} may run on different hardware and will therefore have a different hardware time.
We assume that the control plane is part of a time-synchronized network, e.g., the control plane is synchronized using a protocol such as \acs{PTP}.
Talkers and the control plane in the \acs{TSN} network are synchronized to the network time, but the data plane is not.
This is illustrated in \fig{pdfs/clock_drift-problem}.

\figeps[.4\textwidth]{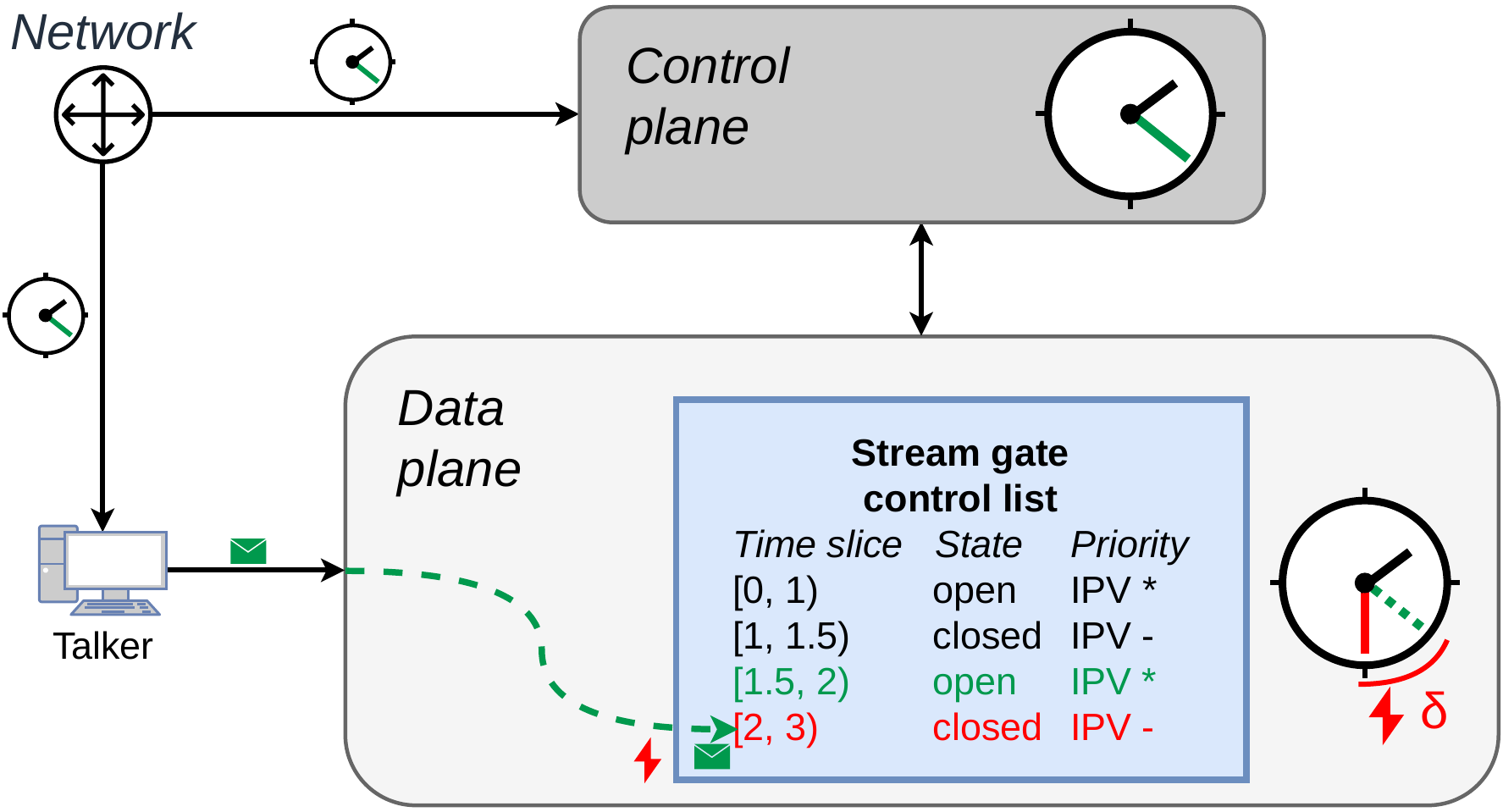}{The control plane and the talker are synchronized to the network time, but the data plane is not. The time differs by the value $\delta$.} 
 
\FI{In \fig{pdfs/clock_drift-problem}}, the data plane has an offset of $\delta$ \FI{relative to} the network time. 
\FI{T}his \FI{offset} results in the network time currently being in the green time slice while the data plane time is in the red time slice.
A frame that should arrive in an open state time slice \FI{may} arrive in a closed state time slice due to lack of time synchronization.
\FI{The offset must be corrected so that the data plane correctly assigns incoming frames to time slices in the \FIi{stream \acs{GCL}} according to the network time.}

\FIi{The second root cause lies within small time delays $\varepsilon_1$ originating from clock drifts or inaccuracies.
Craciunas \textit{et al.} \cite{CrOl21} identify the problem of a temporary loss of time synchronization for time-triggered schedules, leading to clock drift.}
\reviewerii{Clock drifts in P4-PSFP can occur due to delays or inaccuracies in the reception or the generation of hyperperiod packets.
They may be \FI{very small, i.e., in the order of nanoseconds,} but can accumulate over time. 
The value of $\varepsilon_1$ is determined by comparing the measured timestamp $t_j^h$ with the expected duration of $j-1$ hyperperiods since the first completed hyperperiod $t_1^h$.
This is shown in \equa{epsilon1_mod}.
Since $\varepsilon_1$ is always less than or equal to the duration\footnote{If $\varepsilon_1$ is larger than $h$, it simply skips over a complete hyperperiod.} $h$, we can apply the modulo operation in \equa{epsilon1_mod} to simplify the formula to \equa{epsilon1}.
\begin{align}
    \varepsilon_1 &= (\underbrace{(t_j^h - t_1^h)}_{\text{measured}} - \underbrace{(j-1) \cdot h}_{\text{expected}}) \text{ mod } h\label{eq:epsilon1_mod}\\
    \Leftrightarrow \varepsilon_1 &= (t_j^h - t_1^h) \text{ mod } h\label{eq:epsilon1}
\end{align}
}


\FIi{Finally, the third root cause for time inaccuracy results from the configuration of the packet generator in P4-PSFP.
The packet generator in P4-PSFP is configured for one \FIii{ingress} port at a time.
This results in the hyperperiod packet being generated with a small delay $\varepsilon_2$ in between the start of each periodic trigger for each port.
Stream \acp{GCL} on different ingress ports must therefore be synchronized with each other.
This is illustrated in \fig{pdfs/clock_drift-problem_2}.}

\figeps[.4\textwidth]{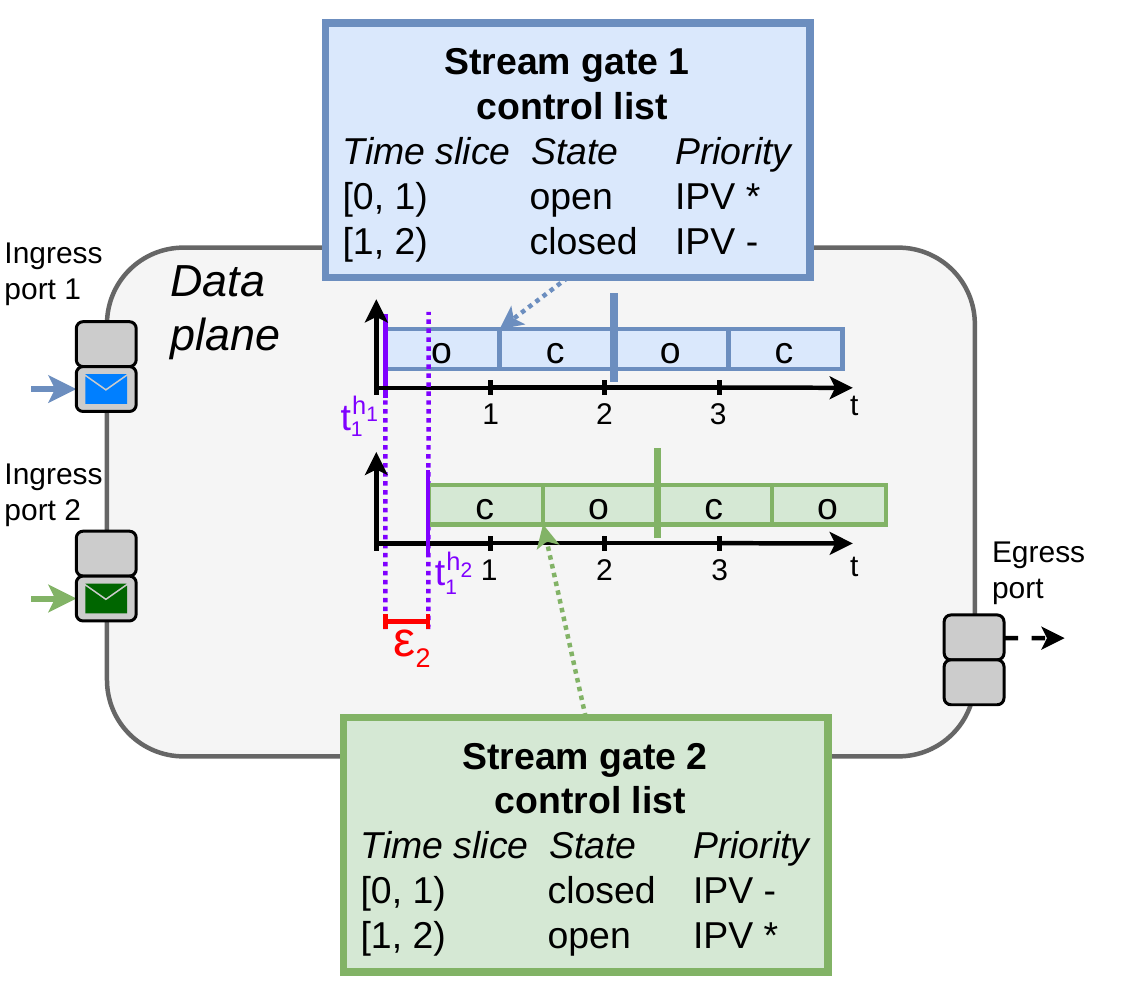}{Two stream \acp{GCL} with complementary gate states are configured on two different ingress ports. Therefore they do not share the same hyperperiodic trigger. The second stream \acs{GCL} starts with a small delay of $\varepsilon_2$.}

\FIi{Two streams on different ingress ports arrive at the P4-PSFP switch in \fig{pdfs/clock_drift-problem_2}.
The stream \acp{GCL} of both streams are configured to have complementary gate states, i.e., only one stream at a time is allowed to transmit.
As both streams are physically on two different ingress ports, they do not share the same hyperperiodic packet generator trigger \FIii{even though they may have the same hyperperiod duration.}
\FIii{The hyperperiodic timestamp $t_j^h$ is required to calculate the relative temporal position $t_i^{rel}$ of a frame $i$, as described in \equa{t_rel}.}
Let $t_1^{h_k}$ be the arrival of the first hyperperiod packet on port $k$.
The delay $\varepsilon_2$ between the two stream \acp{GCL} can then be calculated as the difference between the first hyperperiodic timestamps on each port.
This is shown in \equa{epsilon2}.
\begin{align}
 \varepsilon_2 = t_1^{h_2} - t_1^{h_1} \label{eq:epsilon2}
\end{align}
Due to this delay, the time slices of both stream \acp{GCL} partially overlap by $\varepsilon_2$ in the open state and do not allow for exclusive transmission of one stream at a time.}

\subsubsection{Time Synchronization by $\Delta$-Adjustment}
\label{sec:clock_drift}
In the following, we present a mechanism to adjust the calculated relative position in the stream \acs{GCL} by a dynamic offset value $\Delta = \delta + \varepsilon_1 + \varepsilon_2$.
\FIi{With the proposed approach, P4-PSFP is able to synchronize the data plane with a time-synchronized control plane, account for clock drifts during the runtime, and synchronize stream \acp{GCL} on different ingress ports.}
The control plane measures the delay by reading the hyperperiodic timestamp registers from the data plane and calculates $\varepsilon_1$ and $\varepsilon_2$ according to \equa{epsilon1} and \equa{epsilon2}.
\FIi{The sum of all sources of time inaccuracy $\Delta \in \mathbb{Z}$ is then written by the control plane to a \acs{MAT} in the data plane.
The data plane modifies the relative temporal position $t^{rel}_i$ of a frame in a stream \acs{GCL} according to $\Delta$.}
The mechanism is illustrated in \fig{pdfs/clock_drift-offset} \FI{and further explained below.}

\figeps[.4\textwidth]{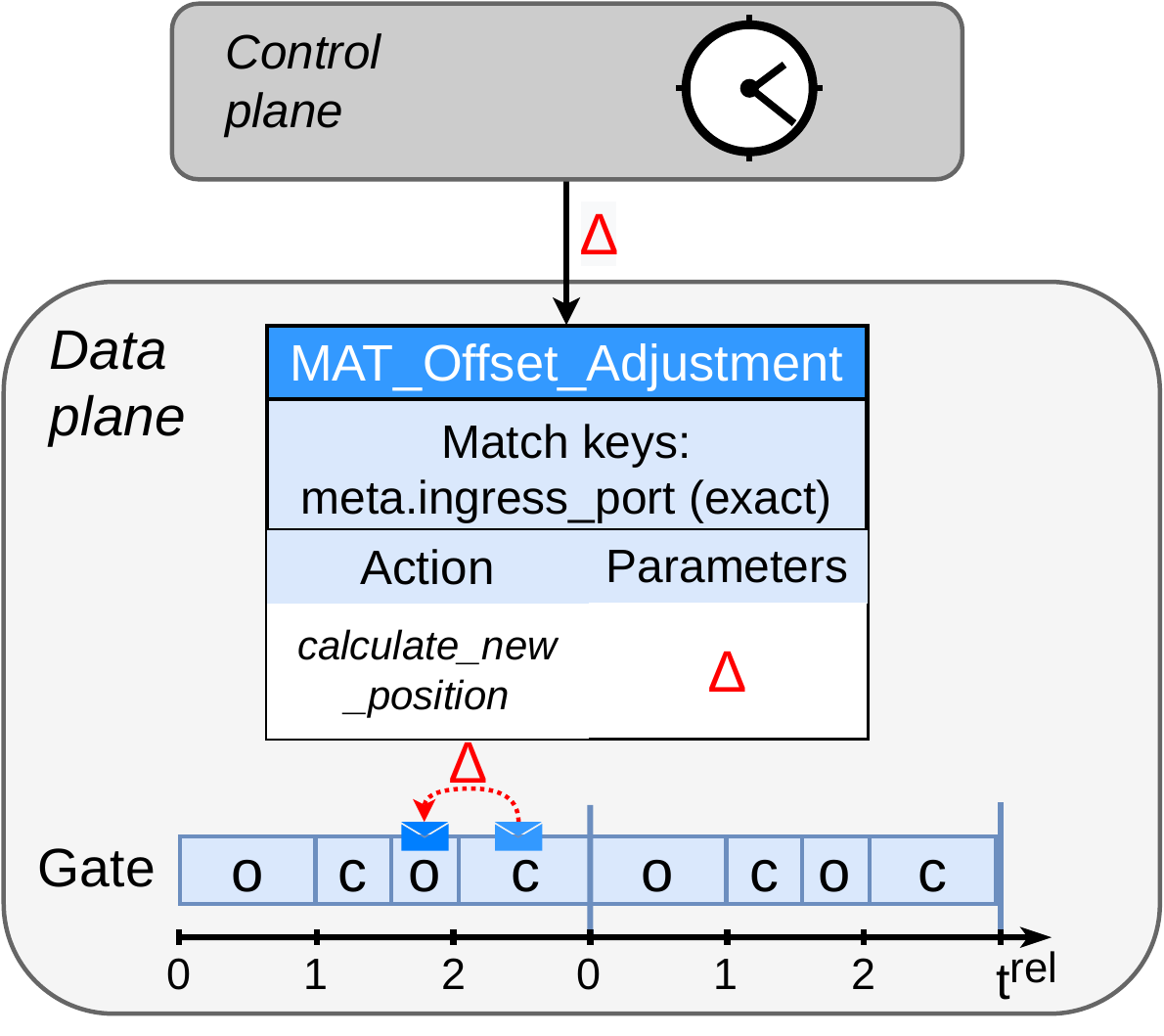}{Due to time inaccuracy, a frame was \FI{erroneously} assigned to the fourth time slice in the \FIi{stream \acs{GCL}}. In fact, this frame arrived in the third time slice of the \FIi{stream \acs{GCL}}. The error is \FI{corrected by subtracting an offset value $\Delta$ from the previously calculated relative position $t^{rel}_i$}.}

\FIi{First, the control plane determines the dynamic offset value $\Delta = \delta + \varepsilon_1 + \varepsilon_2$ as described in \sect{clock_dift_problem}.
\reviewerii{The value of $\varepsilon_1$ is updated every $100 \text{ ms}$} \reviewerii{ to account for varying time inaccuracies and clock drifts during runtime.}
The sum of all individual sources of time inaccuracy $\Delta$ \FIii{per ingress port} is written to a \acs{MAT}.
\FIii{It is important to keep this \acs{MAT} update atomic to not introduce inconsistencies due to multiple \acs{MAT} updates in multiple operations.}
Once this value is written to the \acs{MAT}, the $\Delta$-adjustment is performed entirely in the data plane and adds no further delay to packet processing.}

In the data plane, the action associated with the $\Delta$-adjustment \acs{MAT} modifies the relative position $t^{rel}_i$ of frame $i$.
The value of $\Delta$ is added to the previously calculated relative position $t^{rel}_i$ described in \sect{periodicity}.
The $\Delta$-adjusted relative position of a frame $i$ in a \FIi{stream \acs{GCL}} is determined in \equa{t_rel_delta}.
\begin{align}
    t_{i_\Delta}^{rel} = t^{rel}_i + \Delta \label{eq:t_rel_delta}
\end{align}
By adding the offset value $\Delta$ to the relative position $t^{rel}_i$, the inaccuracy in the relative position of the frame is corrected. The frame may be assigned to a different time slice in the \FIi{stream \acs{GCL}} than it was before.

Altering the relative position and assigning a different time slice \FI{introduces} a new problem.
$\Delta$ is added as an absolute value to the relative position $t^{rel}_i$ of the frame.
Consequently, the $\Delta$-adjustment action may assign a position in the stream \acs{GCL} that does not exist.
The new temporal position may result in an underflow or overflow of a frame relative to its hyperperiod and the packet is erroneously dropped.
\FI{For example, a frame arriving at $t^{rel}_i = 0$ with an offset value of $\Delta = -1$ is assigned to a non-existent negative time slice $t^{rel}_{i_\Delta} = -1$ according to \equa{t_rel_delta}.}
P4-PSFP must \FI{ensure} that the periodicity is maintained even if arbitrary values are added to or subtracted from the relative position without disrupting correct packet forwarding at any time.
A solution to this problem, i.e., the calculation of the relative position \FI{$t^{rel}_{i_\Delta}$ with respect to underflows and overflows}, is illustrated in \fig{pdfs/offset_detection}.

\figeps[0.4\textwidth]{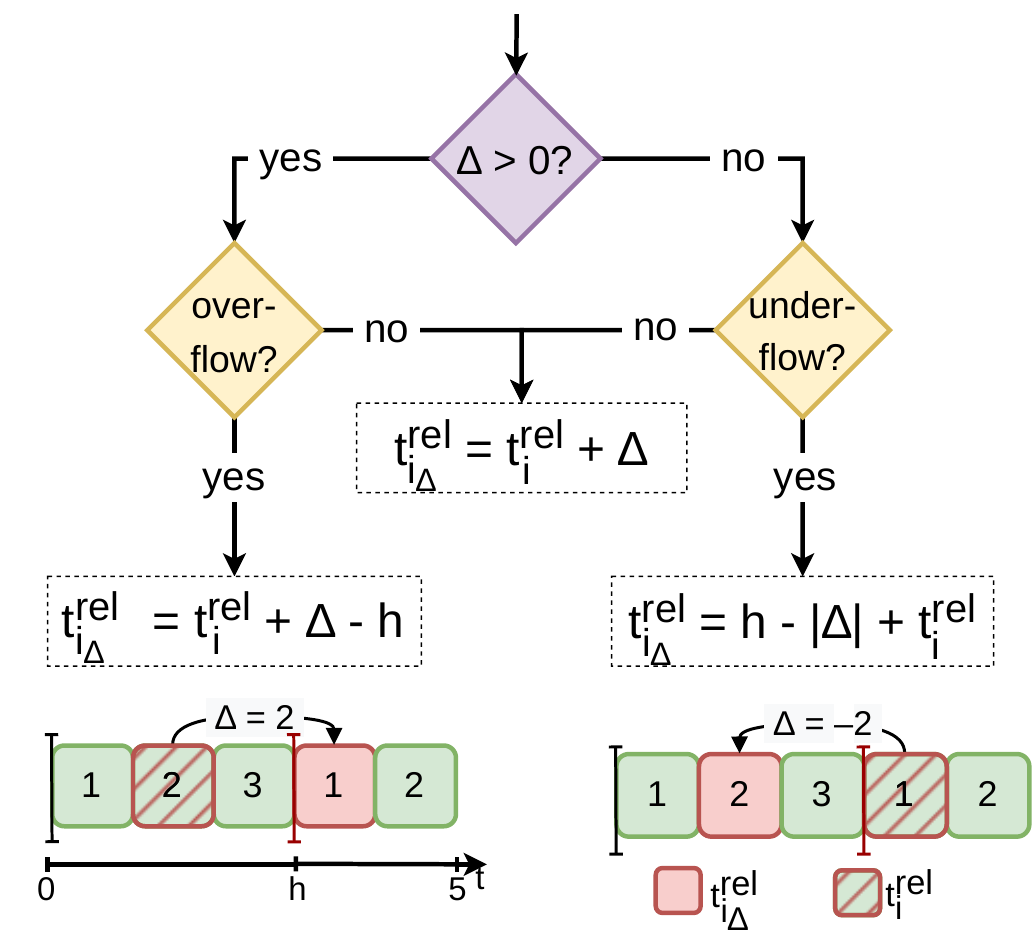}{Possible new positions in the hyperperiod for frames with a hyperperiod of the duration $h = 3$ time steps and an offset $\Delta = \pm 2$ time steps. Simply adding the offset $\Delta = -2$ to the relative timestamp $t^{rel}_i = 1$ results in a new relative position $t^{rel}_{i_\Delta} = -1$, which is not in a valid time slice. P4-PSFP has to account for an under- or overflow with the provided formulas.}

\reviewerii{\fig{pdfs/offset_detection} shows the $\Delta$-adjustment mechanism in the data plane with respect to over- and underflows of the relative position of a frame.}
First, we have to detect whether the offset $\Delta$ is positive or negative, i.e, if $\Delta$ needs to be added or subtracted.
\reviewerii{If it is positive, $t_{i_\Delta}^{rel}$ can potentially overflow, or underflow if it is negative.}
As the data type of $\Delta$ is unsigned, we cannot check for negative values in the data plane.
\FIii{The sign of $\Delta$ is identified by the control plane and written into a \acs{MAT} with a single entry per port \FI{in an atomic operation}
\footnote{By having one \acs{MAT} for positive values and one \acs{MAT} for negative values, the order of updates of the sign \acs{MAT} and of the $\Delta$-\acp{MAT} ensures that no inconsistency is introduced by performing the two \acs{MAT} updates.}.}
\FIii{Letting the control plane handle the sign determination also saves computational resources in the data plane}.

Second, we have to detect an underflow or an overflow.
\reviewerii{On an overflow, the relative position corrected by $\Delta$ will exceed the hyperperiod and be assigned to a non-existent time slice larger than the hyperperiod.
Similarly, on an underflow, the relative position corrected by $\Delta$ is assigned to a non-existent time slice smaller than zero.}
An underflow or an overflow is detected if the newly calculated position $t^{rel}_{i_\Delta}$ results in a very large value, i.e., larger than a hyperperiod or the offset.
\reviewerii{If no over- or underflow is detected, $t_{i_\Delta}^{rel}$ can be calculated according to \equa{t_rel_delta}.}

\reviewerii{Third, if an over- or underflow is detected, we have to adjust the relative position $t^{rel}_i$ to keep its position in a hyperperiod.
To that end, \equa{overflow_delta} must be used on an overflow and \equa{underflow_delta} must be used on an underflow.
\begin{align}
    t_{i_\Delta}^{rel} &= t_i^{rel} + \Delta - h\label{eq:overflow_delta}\\
    t_{i_\Delta}^{rel} &= h - |\Delta| + t_i^{rel}\label{eq:underflow_delta}
\end{align}
For an overflow, it holds that $t_i^{rel} + \Delta > h$.
Therefore, in \equa{overflow_delta} the duration of a full hyperperiod $h$ is subtracted to keep the relative position in the range from $0$ to $h$.
Similarly, for an underflow, it holds that $t_i^{rel} - \Delta < 0$.
Therefore, \equa{underflow_delta} takes a full hyperperiod, subtracts the absolute value of $\Delta$, and adds the relative position $t_i^{rel}$ to stay in the range from $0$ to $h$.}

The \FI{explained approach for time synchronization by $\Delta$-adjustment} is performed in step \ballnumber{2} in \fig{pdfs/implementation}\FI{, i.e, in egress processing before the recirculation}.

%% file: chapters/06-evaluation.tex
\section{Evaluation of P4-PSFP}
\label{sec:eval}
In this section, we evaluate \FI{the} implementation of PSFP in P4 on the Intel Tofino™ switching ASIC.
First, we describe the testbed environment used for our evaluation.
Second, we test the functionality of \FI{flow meters}, \FIi{stream \acp{GCL}}, and the $\Delta$-adjustment.
Third, we evaluate the performance of \FI{the} implementation in a\FI{n} overloaded network environment with and without PSFP enabled and show how it affects the latency.
Finally, we analyze the scalability of \FI{the} implementation on hardware with respect to the number of supported streams when using different stream identification functions.

\subsection{Testbed Environment}
The testbed environment used for our evaluation consists of two switches, one running the traffic generator P4TG\cite{p4tg} and the other running \FI{the} implementation of P4-PSFP.
The environment is illustrated in \fig{pdfs/setting_p4tg}.

\figeps[\columnwidth]{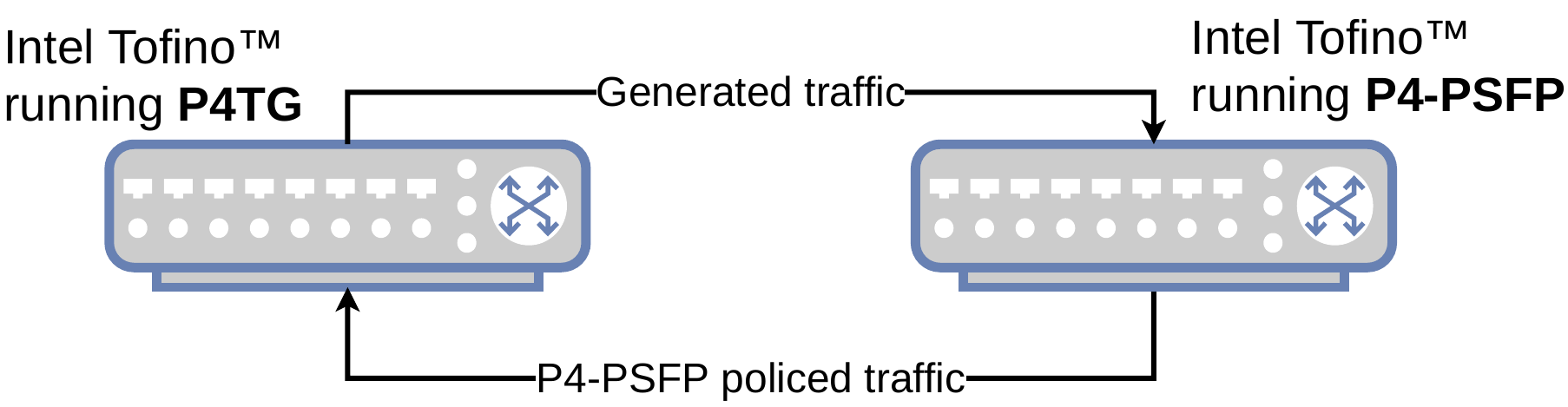}{The testbed environment used for evaluation. Traffic generated by an Intel Tofino™ running P4TG is sent through an Intel Tofino™ running our P4-PSFP implementation and back to P4TG.}

An Edgecore Wedge 100BF-32X switch with an Intel Tofino™ ASIC running \FI{the} P4-PSFP implementation is connected via \SI{100}{\gbps} links to another switch with an Intel Tofino™ ASIC running the P4TG\cite{p4tg} traffic generator.
P4TG generates \acf{CBR} traffic at a transmission rate of \SI{100}{\gbps} and a frame size of 1280 bytes.
\FI{As frames generated by P4TG do not contain a VLAN tag, } we prepend an additional VLAN header to the frame before processing the PSFP pipeline.
This header is removed when the frame leaves the P4-PSFP switch.
The P4-PSFP policed traffic is fed back to the input ports of the P4TG switch to perform latency analysis.
\subsection{Functionality of PSFP Components}
This section presents the experiments to test the functionality of the \acs{PSFP} components in \FI{the} implementation.
We perform tests to verify the functionality of \FI{flow meters, i.e., the credit-based metering, stream gates, i.e., the periodicity of \FIi{stream \acp{GCL}} and the time-based metering, and \FI{the} approach for time synchronization with $\Delta$-adjustment.}
Stream filters and their stream identification are implicitly verified by the following experiments as stream gate and flow meter instances \FI{are} not applied if a stream filter has not identified them. 
We evaluated the implemented PSFP parameters \textit{MaxSDUExceeded}, \textit{GateClosedDueToInvalidRX}, \FI{\textit{GateClosedDueToOctetsExceeded}}, \textit{MarkAllFramesRed}, and \textit{DropOnYellow} \FI{as described in \sect{psfp_summary}}.
However, for the sake of clarity, only the latter two are described in \FI{the} evaluation.
\subsubsection{Functionality of the Flow Meter}
\label{sec:eval_flow_meter}
In this section, the functionality of the flow meter instance, i.e., the credit-based metering and its additional parameters as summarized in \tabl{psfp_params} are evaluated.

The traffic generator P4TG\cite{p4tg} is used to generate a \SI{100}{\gbps} stream which passes through the PSFP mechanism in \FI{the} implementation.
\FIi{The P4-PSFP switch applies the flow meter to the generated traffic stream \FIi{to label frames in their respective colors}.
The control plane measures the traffic rates for each frame color.}

We set the \FI{\acf{CIR}} to \SI{70}{\gbps} and the \FI{\acf{EIR}} to \SI{20}{\gbps}.
The stream gate remains in a permanently open state\FIi{, i.e., no time-based metering is applied.}

\FI{We test \FI{the} implementation \FI{in} three settings \FIi{which are all verified in a single experiment over time.}
First, no additional parameters are set to verify the functionality of the token bucket algorithm and the frame coloring.
Second, the \textit{DropOnYellow} parameter is set which additionally marks yellow-labeled frames red and drops them.
Third, the \textit{MarkAllFramesRed} parameter is set which permanently closes the flow meter instance if a single frame exceeds the \acs{EIR}, i.e., \FIi{once a single frame} is colored red.}

\FI{The experiment runs for \SI{20}{\second}.
Up to $t_0 =$ \SI{2000}{\micro\second}, no additional parameters are set.}
At $t_0$, we set the parameter \textit{DropOnYellow} and at $t_1 = $ \SI{4000}{\micro\second}, we set the parameter \textit{MarkAllFramesRed}.
\FI{\fig{plots/flow_meter_large_p4tg} compiles the \FIi{measured traffic rates} per frame color over time \reviewerii{for the first} \SI{6}{\milli\second} in the experiment. \reviewerii{After} \SI{6}{\milli\second}\reviewerii{, the measured traffic rates do not change anymore.}}

\figeps[\columnwidth]{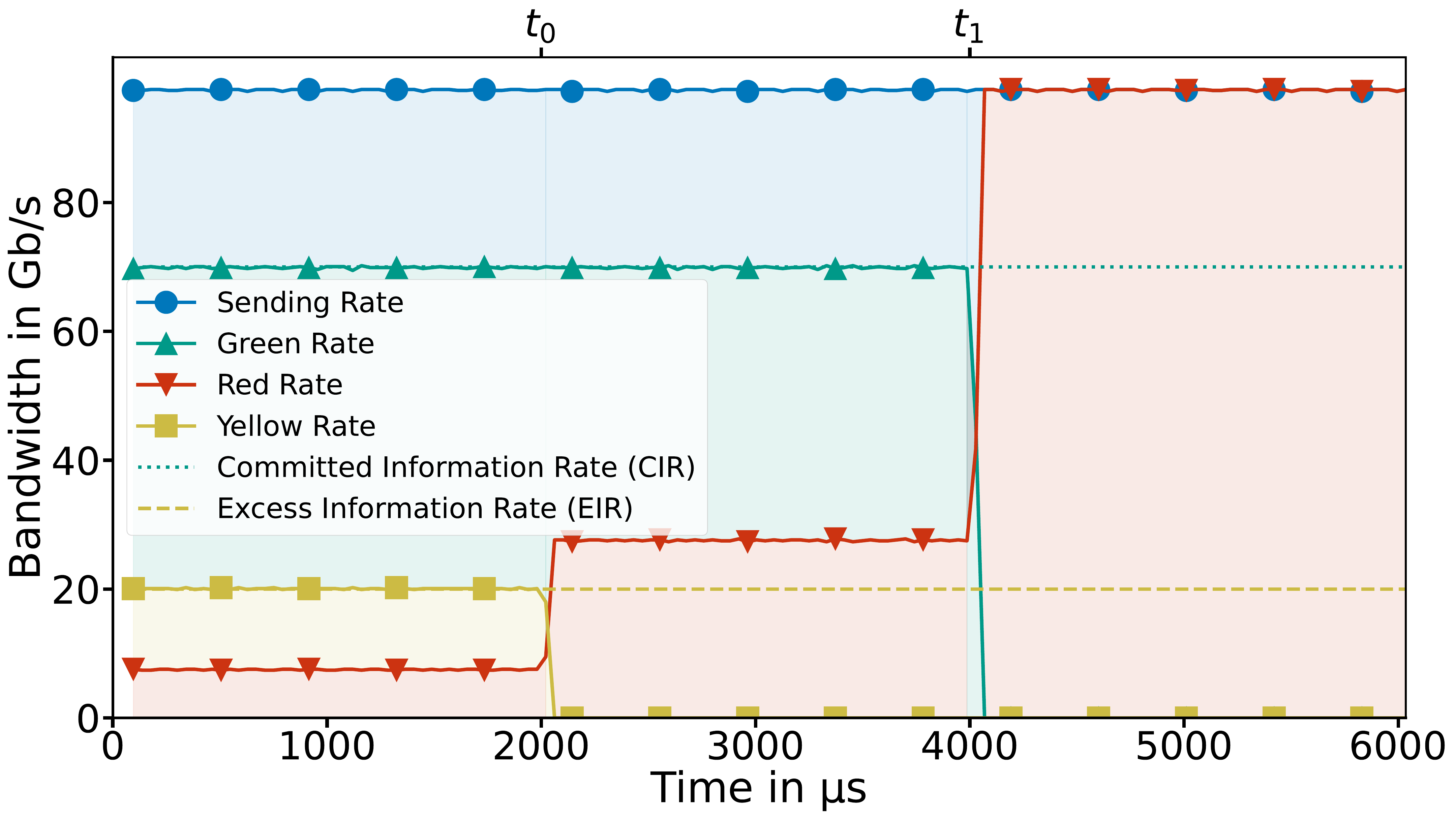}{Rates of colored frames in the flow meter instance while receiving frames at a rate of \SI{100}{\gbps}. The horizontal dashed lines show the configured \acs{CIR} and \acs{EIR}. At $t_0$, the parameter \textit{DropOnYellow} is set and at $t_1$ the parameter \textit{MarkAllFramesRed} is set.}

\begin{figure*}[htb!]
\begin{center}
\leavevmode
  \parbox[t]{1\textwidth}{%
    \resizebox{1\textwidth}{!}{\includegraphics{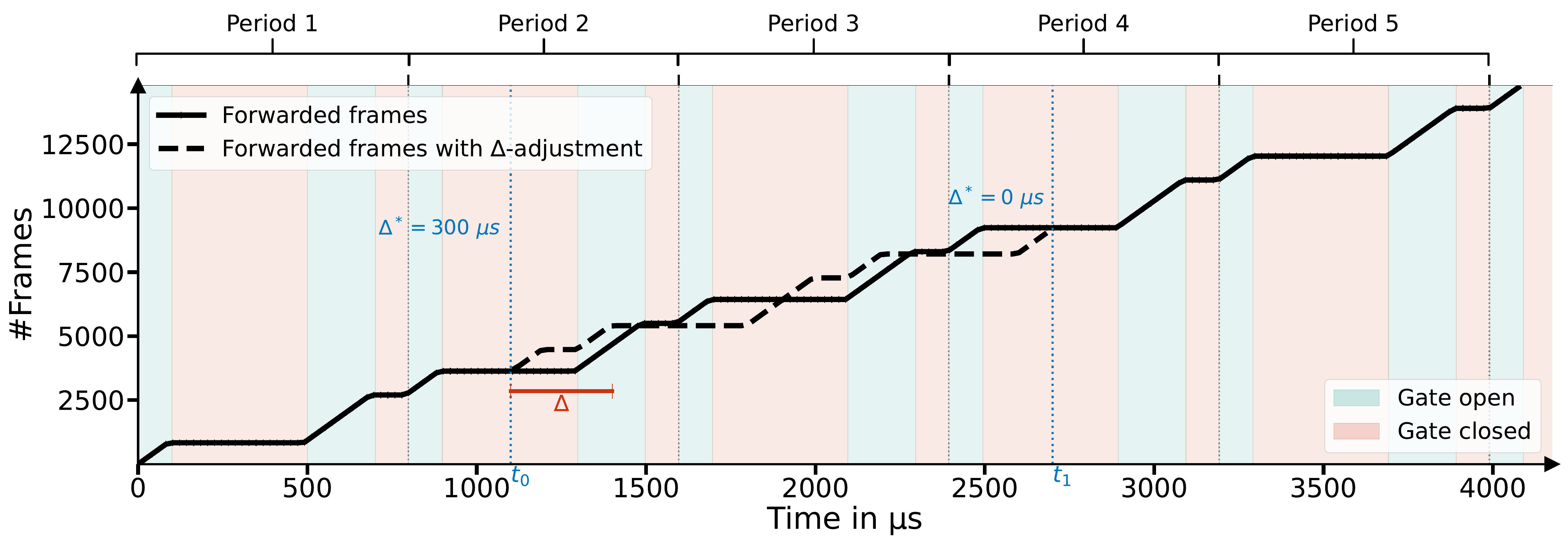}}
  }
\caption{Plot of forwarded frames in the stream gate instance. Periods are indicated by vertical lines. Gate states of the \textit{1-4-2-1} \FIi{stream \acs{GCL}} are illustrated as either red (closed) or green (open). The solid line indicates the forwarded packets without applying the $\Delta$-adjustment. There are five repeating periods of the \textit{1-4-2-1} \FIi{stream \acs{GCL}}. For the dashed line, a $\Delta$-adjustment of $\Delta^* =$ \SI{300}{\micro\second} is applied between $t_0$ and $t_1$.\vspace{-0.2cm}}
\label{fig:plots/schedule_1421_short_time}
\end{center}
\end{figure*}

Up to time $t_0$, the measured \FIi{rates} for the green and yellow labeled frames \FI{align with} the configured rates \FIi{of the \acs{CIR} and \acs{EIR}}.
\FIi{\SI{70}{\gbps} of the total generated traffic stream at a rate of \SI{100}{\gbps} is labeled green while \SI{20}{\gbps} is labeled as yellow. 
The remaining portion of the generated traffic stream is categorized as red.}
The sending rate is slightly below \SI{100}{\gbps} due to the extra bytes added for the recirculation header \FIii{by P4-PSFP as described in \sect{p4_psfp_details}}.

By setting the \textit{DropOnYellow} parameter at $t_0$, in \fig{plots/flow_meter_large_p4tg}, the yellow rate immediately drops to zero while the red rate is increased by the value of the \acs{EIR}.
\FI{Yellow frames are now labeled red and therefore dropped.}

At time $t_1$, we enable the \textit{MarkAllFramesRed} parameter.
As a result, the red rate in \fig{plots/flow_meter_large_p4tg} \FI{aligns} with the sending rate while the green rate drops to zero. 
Frames in this flow meter instance are no longer forwarded.

These experiments show that \FI{the} implementation of the PSFP flow meter instance correctly applies a credit-based metering approach to an incoming stream. 
The parameter \textit{DropOnYellow} successfully drops yellow-labeled frames, and the parameter \textit{MarkAllFramesRed} successfully blocks streams permanently as soon as they \FIi{do not adhere to their \revieweri{stream} descriptor for the first time}.
\subsubsection{Functionality of the Stream Gate}
We define the \textit{1-4-2-1} \FIi{stream \acs{GCL}} to verify the periodicity and the time-based metering in \FI{the} stream gate implementation.
The \textit{1-4-2-1} \FIi{stream \acs{GCL}} consists of four time slices of the duration \SI{100}{\micro\second}, \SI{400}{\micro\second}, \SI{200}{\micro\second} and \SI{100}{\micro\second}. 
\FI{Its period is therefore \SI{800}{\micro\second}.}
The time slice durations are chosen arbitrarily \FI{and have no further meaning}.
\FI{The stream gate states of the \textit{1-4-2-1} \FIi{stream \acs{GCL}} are alternating, starting in the open state.}

Again, the traffic generator P4TG\cite{p4tg} is used to generate a \SI{100}{\gbps} \acs{CBR} traffic stream.
\FIi{The flow meter instance is disabled for the evaluation of the time-based metering, i.e., no credit-based metering is applied in this experiment.}

The objectives of this experiment are twofold.
\FIii{First, the time-based metering and the periodicity of stream \acp{GCL} are verified.
Second, the ability of the $\Delta$-adjustment to apply an offset $\Delta^*$ at any given time without introducing inconsistencies is verified.
For the first objective, we apply the \textit{1-4-2-1} stream \acs{GCL} to the generated traffic stream.
For the second objective, we additionally introduce a $\Delta$-adjustment of $\Delta^* =$ \SI{300}{\micro\second} at $t_0 =$ \SI{1100}{\micro\second}.
At $t_1 =$ \SI{2700}{\micro\second}, we reset the offset $\Delta^*$ to \SI{0}{\micro\second}.
We measure the number of forwarded frames by the stream gate instance in the P4-PSFP switch over time to verify the objectives.}

\FIi{The results are compiled in \fig{plots/schedule_1421_short_time}}. 
In this plot, a rising edge indicates active frame forwarding while a plateau indicates the suspension of frame forwarding.
In addition, the time slices of the \FIi{stream \acs{GCL}} in the closed state are \FIii{highlighted} in red whereas those in the open state are \FIii{marked} in green.
The solid line illustrates five consecutive cycles of the \textit{1-4-2-1} \FIi{stream \acs{GCL}}.
The number of cycles measured during the runtime of \SI{4000}{\micro\second} corresponds to the expected number of cycles of the \FIi{stream \acs{GCL}} with a period of \SI{800}{\micro\second}.
Moreover, the rising edges correspond exactly to the areas marked in green and the plateaus correspond to the areas marked in red, i.e., the four measured alternating time slices per period exactly align with the predefined time slices specified in the \FIi{stream \acs{GCL}}.
This experiment validates the time-based metering, and the proposed solution for achieving the periodicity of \FIi{stream \acp{GCL}} by using the internal traffic generator as described in \sect{periodicity}.

\FIi{The second experiment is represented by the dashed black line in the plot.
In this experiment, we set the offset value $\Delta^* =$ \SI{300}{\micro\second} at $t_0 =$ \SI{1100}{\micro\second}. 
\FIii{Consequently, the curve follows the curve without the $\Delta$-adjustment for \SI{1100}{\micro\second} and is then shifted by $\Delta^*$ to the left during the second period.}
Each frame is now assigned to a time slice determined by its ingress timestamp plus \SI{300}{\micro\second}.
The third period corresponds to a period in which all time slices of the \textit{1-4-2-1} \FIi{stream \acs{GCL}} are shifted by $\Delta^*$.
In the fourth period, we reset the offset value $\Delta^*$ to \SI{0}{\micro\second} at $t_1 =$ \SI{2700}{\micro\second}.
At this point, the curve immediately returns to its original path without the $\Delta^*$ adjustment.
Finally, the fifth period mirrors the characteristics of the first period because no $\Delta^*$ value is applied.
Through this experiment, we demonstrate that the proposed approach to account for time synchronization can seamlessly introduce an arbitrary offset value at an arbitrary time.
This compensates for time inaccuracies such as accumulated clock drift or inaccuracies without causing inconsistencies in forwarding.}

\subsection{\FI{PSFP Application to Competing Streams in an Overloaded Network}}
In this section, we evaluate \FI{the} P4-PSFP implementation in a setting \FI{where two streams compete for limited resources in an overloaded network}. \FI{First, we give an overview. Then we describe \FI{the} setup and finally, we present the results.}
\subsubsection{\FI{Overview}}
In a typical \acs{TSN} setup, \FI{time-triggered talkers communicate in real-time.}
A traffic shaper such as the \acs{TAS} is used to schedule this time-sensitive traffic.
Since the stream gate component of PSFP \FI{can be explained as} a \acs{TAS} not bound to priority queues, we \FI{leverage the} implementation of \FI{P4-PSFP} to protect the schedules of a hypothetical \acs{TAS} by dropping frames received in a disallowed time slice before they enter the \acs{TAS}.
We leverage the traffic generator P4TG to force queueing by overloading the capacity of a \SI{100}{\gbps} link with two \SI{90}{\gbps} streams, thereby introducing queueing and increasing latency.
Higher latency can cause frames to arrive outside their intended time slice in the \acs{TAS} schedule.
\FI{We aim to protect the schedules of the hypothetical \acs{TAS} by applying different configurations of \FIi{stream \acp{GCL}} in PSFP.}
We hypothesize that the latency on a congested network link can be reduced with \FI{P4-PSFP}.
\subsubsection{\FI{Experiment Setup}}
We use the traffic generator P4TG to generate two streams of \SI{90}{\gbps} each.
Each stream has its own ingress and recirculation port in the P4-PSFP switch. 
However, both streams are sent back to the P4TG switch via the same egress port after PSFP is applied.
This will overload the egress network link \FIi{which has a} capacity of \SI{100}{\gbps}.
An overview of \FI{the experiment setup} is given in \fig{pdfs/eval_performance}.

\figeps[\columnwidth]{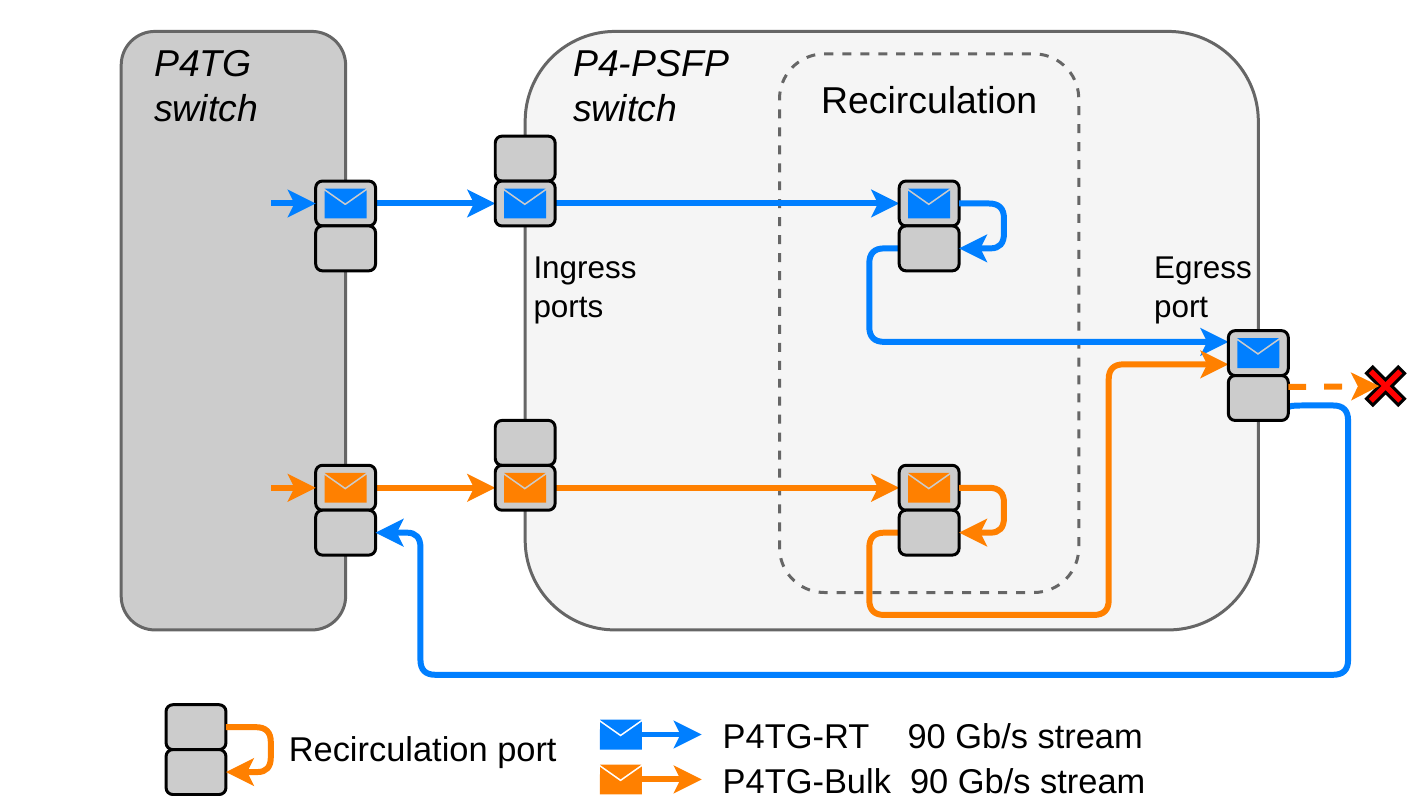}{Two \SI{90}{\gbps} streams are generated by P4TG and sent to the P4-PSFP switch. Each stream has its own ingress and recirculation port, but both streams are sent back through the same egress port \FI{which causes} queueing. Frames from the P4TG-\FI{RT} stream are used to measure the latency while frames from the P4TG-Bulk stream are dropped after queueing.}

\threesubfigeps{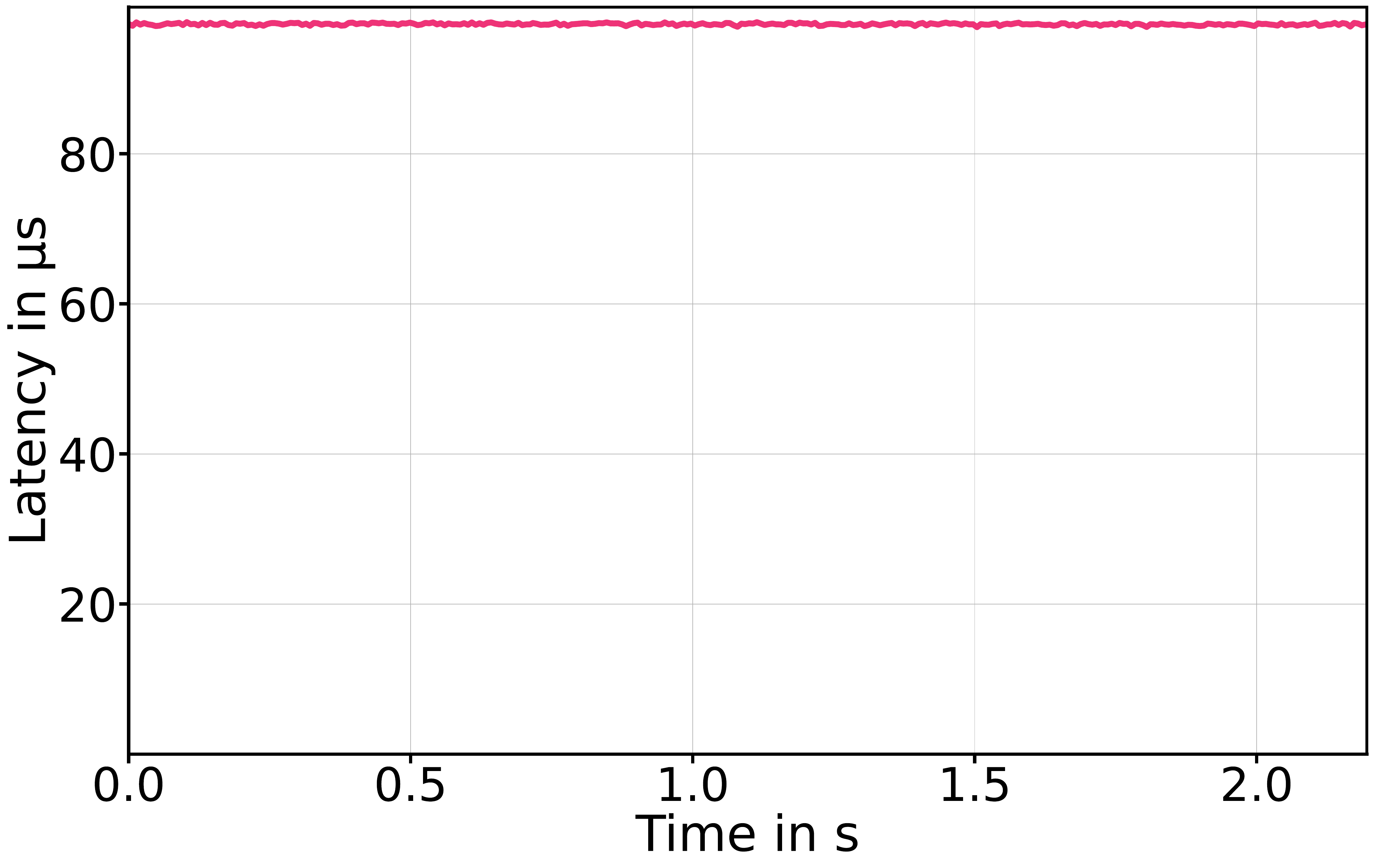}{PSFP disabled.}{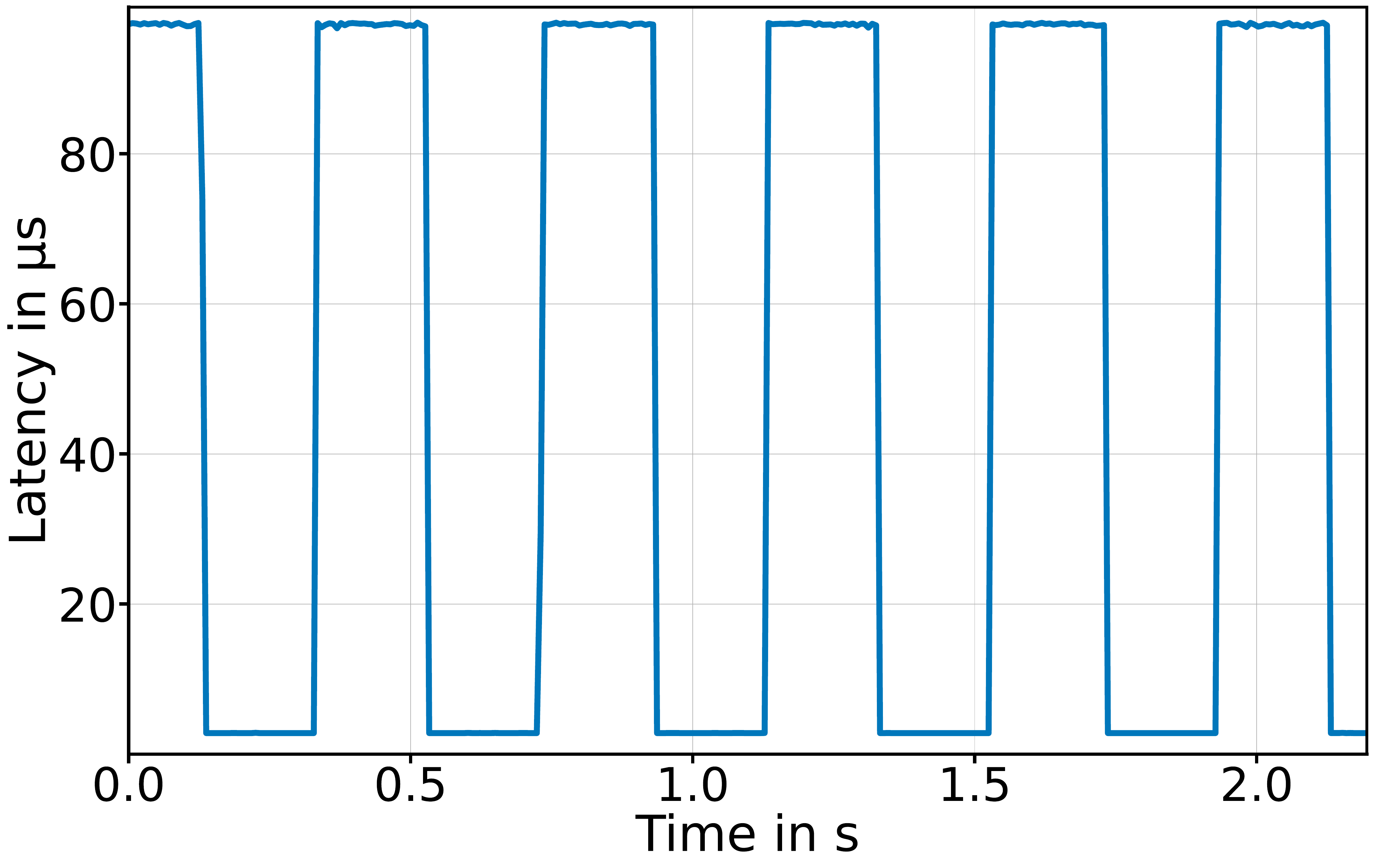}{One stream scheduled with the \textit{50-50} \FIi{stream \acs{GCL}}.}{plots/50_50_rtt}{Both streams scheduled with complementary \textit{50-50} stream \acp{GCL}.}{Latency of the P4TG-\FI{RT} stream in \FI{the} congested network setting after applying different \FIi{stream \acs{GCL}} configurations to both P4TG streams.}

\FI{Two streams \FIi{generated by P4TG} are considered in our experiment setup. 
The first is the P4TG-RT stream which has a strict real-time requirement and a tight latency constraint.
Some industrial control applications require meeting strict latency limits, sometimes as low as a few microseconds, as stated in \cite{NaTh19}.
Second, the P4TG-Bulk stream represents traffic that interferes with the P4TG-RT stream and has no real-time requirement.}

\FI{Since we do not have a TSN testbed environment with synchronized talkers, P4TG} transmits both streams continuously, regardless of the allowed transmission times. 
Frames received during time slices in the closed state are dropped by the stream gate instance in P4-PSFP before egress queuing and therefore do not affect the latency.

Sending two \SI{90}{\gbps} streams through the same egress port will overload the egress network link and cause queueing, which increases the latency.
We \FIi{consider} the latency caused by queueing \FIi{in the egress port of the P4-PSFP switch} by applying \FI{three different \FIi{stream \acp{GCL}}} in \FI{the} implementation.
\FI{First, we apply a \FIi{stream \acs{GCL}} that leaves the stream gates of both streams in a permanently open state to measure the latency in the congested network as a baseline without PSFP.
Second, we apply the \textit{50-50} \FIi{stream \acs{GCL}}, which consists of a \SI{200}{\milli \second} time slice in the closed state, followed by a \SI{200}{\milli \second} time slice in the open state to the P4TG-Bulk stream.}
\FI{The stream gate instance associated with the P4TG-\FI{RT} stream remains permanently open.}
Third, we apply the \textit{50-50} \FIi{stream \acs{GCL}} to the P4TG-Bulk stream and an inverted \textit{50-50} \FIi{stream \acs{GCL}} to the P4TG-\FI{RT} stream, i.e., only one stream is allowed to transmit at a time. 

\FIi{Stream \acp{GCL} configured on different ingress ports are not synchronized with each other because the hyperperiodic packet generator trigger is configured on a per-ingress-port basis, as explained in \sect{clock_dift_problem}.
As a result, time slices from the P4TG-RT stream \acs{GCL} overlap with time slices from the P4TG-Bulk stream \acs{GCL} by an offset $\varepsilon_2$.
To solve this problem, we apply the $\Delta$-adjustment to synchronize the two stream \acp{GCL}.
The difference between the stored hyperperiod timestamps $\varepsilon_2$ of both stream \acp{GCL} is calculated by the control plane at the beginning.}

\FI{The P4-PSFP switch drops frames} from the P4TG-Bulk stream after PSFP processing\FIi{, i.e., after egress queueing,} to measure only the latency of the P4TG-\FI{RT} stream in P4TG.

\subsubsection{Results}
In this section, we present the results of the three experiments with different \FIi{stream \acp{GCL}}.
In \FI{the setting where PSFP is disabled}, the \SI{100}{\gbps} link becomes congested and frames are queued in the egress port, increasing the latency. 
The latency remains consistently high at $\approx$ \SI{98}{\micro\second}. 
This can be seen in \fig{plots/open_open_rtt}.

\FI{In the second setting, the \textit{50-50} \FIi{stream \acs{GCL}} is applied to the P4TG-Bulk stream}.
In this setting, the latency drops to \SI{3}{\micro\second} for \SI{200}{\milli\second} during time slices where the P4TG-Bulk stream is blocked, and rises to \SI{98}{\micro\second} when both streams are transmitting.
This is visible in \fig{plots/open_fifty_rtt}.

\FI{In the third experiment, the stream gate is configured to only allow one stream to transmit at a time}.
Here, the latency drops to \SI{3}{\micro\second} permanently.
This is shown in \fig{plots/50_50_rtt}.
\FIii{The exact alignment of the time slices in both stream \acp{GCL} allows for the exclusive transmission of one stream at a time without inconsistencies during state transitions.
This is achieved by synchronizing the two stream \acp{GCL} on two different ingress ports with the $\Delta$-adjustment.}

These experiments show that \FI{the} implementation of P4-PSFP is capable of protecting \acs{TAS} schedules with time-based metering and appropriate \FIi{stream \acs{GCL}} configuration. 
\FIi{The latency reduction achieved by P4-PSFP effectively shows the elimination of queueing in the overloaded egress port queue.
This is accomplished by configuring P4-PSFP with two stream \acp{GCL} that alternate between the two streams and by applying the $\Delta$-adjustment.
The application of the $\Delta$-adjustment shows that P4-PSFP is able to synchronize multiple stream \acp{GCL} on different ingress ports to each other.
Frames that enter the \acs{TAS} after P4-PSFP processing no longer experience congestion-related delay.}

\subsection{Scalability}
\input{chapters/06-evaluation_scale_table}

In this section, we evaluate the scalability of \FI{the} implementation on the Intel Tofino™.
The resources used during the processing of a frame, i.e., tables, actions, and externs, are limited to maintain the line rate of \SI{100}{\gbps}.
All resources used share a common memory. 
\FIi{Sufficient memory capacity must be available to hold the used resources.}
Larger tables, i.e., tables with more key fields, actions, and action parameters, require more memory.
\FI{The maximum size of \acp{MAT}, i.e. the number of entries a \acs{MAT} can \FIi{hold}, must be defined at compile time and cannot be changed at runtime.
\FIi{If the defined maximum size of a \acs{MAT} exceeds the available resources, a compilation error occurs and the P4 program cannot be executed.
\FIii{An overview of the available \acp{MAT} in P4-PSFP is given in \sect{implementation_overview}.}
The maximum number of entries in the stream gate \acs{MAT} to achieve a successful compilation of P4-PSFP proved to be 2048 entries, i.e., 2048 different time slices of stream \acp{GCL} can be modeled.
We gradually increase the size of the stream identification \acs{MAT} while keeping the stream gate \acs{MAT} size at 2048 entries to achieve a successful compilation of P4-PSFP that finds the maximum possible number of supported streams.}}
\FIi{We set the size of the flow meter \acs{MAT} equal to the size of the stream identification \acs{MAT}.}

Using different stream identification functions, i.e., a different composition of the stream identification \acs{MAT} key, results in a different maximum number of streams that can be supported.
Matching fields for stream identification are either used as \textit{exact} or as \textit{ternary} type.
Ternary matching fields allow for aggregation of streams, or for wildcards, meaning that a single ternary entry in the \acs{MAT} can match multiple different streams.
As a tradeoff, a ternary matching entry \FIi{consumes} more memory than an exact matching entry.
An overview of different stream identification \acs{MAT} compositions and the number of entries supported is compiled in \tabl{stream_id}.

\FIi{P4-PSFP} can support up to 35840 different streams when using 2048 time slices with the mandatory null stream identification function\cite{cb} \FI{that only matches on the Ethernet destination address and the VLAN ID}.
If we use exact IP stream identification, \FI{the number of entries} reduces to 32768 different streams for exact matches, or 2048 entries for aggregated wildcard streams with ternary matches.
\FIi{In the survey conducted by Stüber \textit{et al.} \cite{StOs23}, the number of streams discussed in various research papers reached up to 10812 streams in the case of \cite{VlBr22}.
Consequently, to accommodate the number of streams required for the most extensive TSN configuration described in \cite{StOs23}, we can \FIii{choose} either null stream identification or exact IP stream identification while having a total of 2048 time slices in stream \acp{GCL}.}

%% file: chapters/06-evaluation_scale_table.tex
\begin{table*}[htb!]
	\caption{Overview of different stream identification functions according to IEEE 802.1CB\cite{cb}. Depending on the match fields and types, different amounts of streams can be supported in our implementation of PSFP.}
	\begin{center}\resizebox{.85\textwidth}{!}{
        \begin{tabular}{|l||l|l|l|l|}
        \hline
        \textbf{Stream identification function} & Null stream & Source MAC & IP stream (ternary) & IP stream (exact) \\
        \hline\hline
        \textbf{Ethernet source address} & & \begin{tabular}{ll}\checkmark & ex.\end{tabular} & \begin{tabular}{ll}\checkmark & ter.\end{tabular} & \\
        \hline
        \textbf{Ethernet destination address} & \begin{tabular}{ll}\checkmark & ex.\end{tabular} & \begin{tabular}{ll}\checkmark & ter.\end{tabular} & \begin{tabular}{ll}\checkmark & ter.\end{tabular} & \begin{tabular}{ll}\checkmark & ex.\end{tabular} \\
        \hline
        \textbf{VLAN ID} & \begin{tabular}{ll}\checkmark & ex.\end{tabular} & \begin{tabular}{ll}\checkmark & ex.\end{tabular} & \begin{tabular}{ll}\checkmark & ex.\end{tabular} & \begin{tabular}{ll}\checkmark & ex.\end{tabular} \\
        \hline
        \textbf{IP source address} & & & \begin{tabular}{ll}\checkmark & ter.\end{tabular} & \begin{tabular}{ll}\checkmark & ex.\end{tabular} \\
        \hline
        \textbf{IP destination address} & & & \begin{tabular}{ll}\checkmark & ter.\end{tabular} & \begin{tabular}{ll}\checkmark & ex.\end{tabular} \\
        \hline
        \textbf{DSCP} & & & \begin{tabular}{ll}\checkmark & ter.\end{tabular} & \begin{tabular}{ll}\checkmark & ex.\end{tabular} \\
        \hline
        \textbf{Next Protocol} & & & \begin{tabular}{ll}\checkmark & ter.\end{tabular} & \begin{tabular}{ll}\checkmark & ex.\end{tabular} \\
        \hline
        \textbf{Source port} & & & \begin{tabular}{ll}\checkmark & ter.\end{tabular} & \begin{tabular}{ll}\checkmark & ex.\end{tabular} \\
        \hline
        \textbf{Destination port} & & & \begin{tabular}{ll}\checkmark & ter.\end{tabular} & \begin{tabular}{ll}\checkmark & ex.\end{tabular} \\
        \hline \hline
        \textbf{Max. number of stream identification entries} & 35840 & 4096 & 2048 & 32768 \\
        \hline
        \end{tabular}}
	\label{tab:stream_id}
	\end{center}
\end{table*}

%% file: chapters/065-discussion.tex
\section{Discussion}
\label{sec:discussion}
In this section, we discuss the effect of recirculation on the implementation and requirements to stream \acp{GCL} in P4-PSFP. 
Further, we discuss the requirements for porting P4-PSFP to other hardware targets.
We also describe applications that can take advantage of the \SI{100}{\gbps} bandwidth capability of P4-PSFP.
\subsection{The Effect of Recirculation in P4-PSFP}
 \label{sec:discuss_recirculation}
\reviewerii{
The concept of recirculation allows a packet to be processed multiple times, iteratively manipulating its header data.
A packet that undergoes a recirculation requires capacity in terms of bandwidth, e.g., a packet that is recirculated once doubles its total required bandwidth.
For this purpose, the Intel Tofino™ has internal ports that provide capacity for recirculation.
Physical ports can be configured to be used as recirculation ports to increase the capacity available for recirculation.
If the recirculation capacity is exceeded, e.g., by directing more than \SI{100}{\gbps} of traffic to a \SI{100}{\gbps} port, queueing will occur and packets may be dropped.
Furthermore, a recirculation on the Intel Tofino™ adds a constant amount of time\footnote{If the recirculation port is not overloaded.} to the processing delay.}

\reviewerii{In P4-PSFP, one recirculation is required for each frame to retrieve the frame size for the stream filter instance as described in \sect{max_filter}.
Each of the eight ingress ports in P4-PSFP has a dedicated recirculation port to prevent exceeding the available recirculation capacity.
If the ingress ports operate at less than \SI{100}{\gbps}, e.g., at \SI{10}{\gbps}, a single recirculation port is sufficient for the P4-PSFP switch.
On the Intel Tofino™, each recirculation port experiences the same constant sub-microsecond delay.
Therefore, the recirculation delay in P4-PSFP can be considered part of the overall processing delay of each frame imposed by the hardware device.
For matching the frame to its time slice in the stream \acs{GCL}, the ingress timestamp before a recirculation is used.
This allows a frame to be assigned to a time slice according to its arrival time at the switch.}
\subsection{Requirements in P4-PSFP}
In this section, we summarize and discuss the requirements to stream \acp{GCL} in P4-PSFP.
We further discuss the requirements of other P4-based hardware targets to support P4-PSFP.
\revieweri{
\subsubsection{Requirements to Stream GCLs}
In P4-PSFP, all individual stream \acp{GCL} on the same ingress port must be extended to a hyperperiod which is formed by the \acs{LCM} of the periods of all stream \acp{GCL}.
Forming a hyperperiod to model the individual stream \acp{GCL} is a common practice according to Stüber\textit{ et al.}\cite{StOs23} and is not specific to P4-PSFP. 
Specifically for P4-PSFP, a hyperperiod must be in the range of \SI{2}{\micro\second} to $\approx$ \SI{2.1}{\second}. 
According to a survey by Stüber \textit{et al.}\cite{StOs23}, this is not a limitation as stream hyperperiods in their survey range from \SI{32}{\micro\second} to \SI{500}{\milli\second}.}

\subsubsection{Requirements to Other Hardware Targets}
\reviewerii{For P4-PSFP to be portable to other hardware targets, the components of the \acs{PSFP} mechanism, namely the stream filter, the stream gate, and the flow meter, must be implemented.
While the stream filter component is mostly covered by basic \acp{MAT}, the flow meter component must be implemented in the form of the token bucket algorithm, e.g., by a meter extern.
For the stream gate component, the main requirement is a trigger to determine the end of a hyperperiod.
This trigger must indicate the end of the hyperperiod with a hardware-based timestamp.
On the Intel Tofino™, the internal traffic generator is used for this purpose.
On other hardware, a CPU-based generator could be leveraged.
Inaccuracies caused by the hyperperiod trigger can be compensated by the mechanism presented in \sect{clock_drift}.}

\reviewerii{Another requirement for the implementation is a strong time synchronization.
Ideally, the data plane is synchronized via PTP.
Otherwise, the control plane can be synchronized with PTP and the $\Delta$-adjustment mechanism can be used to account for the synchronization in the data plane.}

\subsection{P4-PSFP across Other Domains}
\revieweri{
The provided P4-PSFP implementation supports rates up to \SI{100}{\gbps} while most \acs{TSN} networks operate with line rates of at most \SI{1}{\gbps}.
DetNet extends the reach of \acs{TSN} networks into the IP and MPLS domain as described in \sect{detnet}.
The DetNet working group focuses on networks that are under a single administrative control, such as private WANs or campus-wide networks.
In this environment, \acs{TSN} sub-networks could operate at higher bandwidths and benefit from the \acs{PSFP} mechanism.}

\revieweri{Furthermore, the IEEE 802.3ch working group is currently discussing a standard for single-pair Ethernet connections for automotive applications with up to \SI{10}{\gbps}\cite{ch}.
Since \acs{TSN} mechanisms are already deployed in automotive applications, such a standard would benefit from a \acs{PSFP} implementation capable of higher bandwidths.}

\revieweri{Another emerging area is distributed AI in data centers. 
Here, network latency has become increasingly important, and methods such as Remote Direct Memory Access (RDMA) are being used to drastically reduce latency with a shared memory.
Collective AI operations distribute and collect their data over the network to enable for interaction between multiple compute nodes.
This requires high bandwidth and low latency.
While this is not a typical application for \acs{TSN} networks, the policing of the \acs{PSFP} mechanism could help reduce latency in such data centers.
However, PSFP also requires a highly synchronized network and scheduled traffic, which is usually not the case in such environments.
Applying the concept of PSFP to data-center networks is an interesting concept that is left open for future work.}


%% file: chapters/07-conclusion.tex
\section{Conclusion}
\label{sec:conclusion}
This paper presents P4-PSFP, a novel implementation of the PSFP mechanism in the P4 language on a hardware-based switching ASIC.
\FIi{P4-PSFP} is, to the best of our knowledge, the first full-fledged implementation of PSFP on real hardware \FI{that supports the full set of functions conform to IEEE Std 802.1Qci\cite{qci}}.
We successfully identified and solved challenges encountered when using real hardware, such as constrained computational resources, achieving periodicity of stream \acp{GCL}, \FIi{and time synchronization}.

\FI{T}he functionality of each of the PSFP components, i.e., stream filters, stream gates, and flow meters \FI{was verified in extensive evaluations, including the correct operation of the credit-based metering, stream \acp{GCL} and their periodicity, i.e., the time-based metering, and the $\Delta$-adjustment for time synchronization.}

Furthermore, \FI{P4-PSFP} was evaluated in a network environment where two links with an overall transmitting rate of \SI{180}{\gbps} overload a link with a capacity of \SI{100}{\gbps}, \FIi{causing queueing in the egress port queue.}
By configuring stream \acp{GCL} to allow only one stream transmission at a time, \FIi{\FI{P4-PSFP}
effectively eliminates queueing in the congested egress port queue through highly accurate synchronization of the stream \acp{GCL}.}

\FI{The scalability analysis showed that P4-PSFP can support 2048 different stream \acs{GCL} time slices, and up to 35840 different streams, depending on the stream identification function. The code of P4-PSFP is available on GitHub\cite{git}}.

\FIi{P4-PSFP can be used to ensure that streams in a TSN network conform to their announced \revieweri{stream} descriptors until a full-fledged hardware implementation on appropriate TSN switches is available.}
\FIii{Moreover, the concepts implemented in P4-PSFP can be individually leveraged and reused in other implementations.}
The proposed approach to account for time inaccuracy of time-critical components in the data plane, e.g., stream \acp{GCL}, can be utilized in various implementations where highly accurate time synchronization is required and protocols such as \acs{PTP} are not available.
The mechanism described for achieving the periodicity of schedules by using the internal traffic generator of the Intel Tofino™ can be used in other implementations that require periodic behavior.

%% file: glossar.tex
\section*{List of abbreviations}
\begin{acronym}
    \acro{QoS}{quality of service}
	\acro{TSN}{Time-Sensitive Networking}
    \acro{GCL}{gate control list}
	\acro{PTP}{Precision Time Protocol}
    \acro{IAT}{inter-arrival time}
    \acro{DPTP}{Data Plane Time Synchronization Protocol}
	\acro{PSFP}{Per-Stream Filtering and Policing}
	\acro{IPV}{Internal Priority Value}
	\acro{PCP}{Priority Code Point}
	\acro{CIR}{Committed Information Rate}
	\acro{EIR}{Excess Information Rate}
	\acro{DEI}{DropEligibleIndicator}
	\acro{P4}{Programming Protocol-independent Packet Processors}
	\acro{MAT}{match+action table}
	\acro{LPM}{longest-prefix-match}
	\acro{PSA}{Portable Switch Architecture}
	\acro{TNA}{Tofino Native Architecture}
     \acro{TDMA}{Time Division Multiple Access}
     \acro{LCM}{least common multiple}
     \acro{TAS}{Time-Aware Shaper}
     \acro{CBS}{Credit-Based Shaper}
     \acro{CBR}{constant bit-rate}
\end{acronym}